%% file: main.tex
\documentclass[
  journal=largetwo,
  manuscript=article-type,
  year=2020,
  volume=37
]{cup-journal}

\usepackage{aas-macros}
\usepackage[utf8]{inputenc}
\usepackage{csquotes}
\usepackage[english]{babel}

\usepackage{amsmath,amssymb}
\usepackage[nopatch]{microtype}
\usepackage{booktabs}
\usepackage{siunitx}
\usepackage{graphicx}
\usepackage{rotating}
\usepackage{multirow}
\usepackage{tablefootnote}
\usepackage{tikz}
\usepackage{upgreek}
\usepackage{soul}

\DeclareNameAlias{sortname}{family-given}
\newcommand{\DMeg}{\ensuremath{{\rm DM}_{\rm EG}}}

\newcommand{\zdm}{{\sc zDM}}
\newcommand{\pccc}{\si{pc\,\centi\metre^{-3}}}

\newcommand{\halflight}{\ensuremath{\phi}}
\newcommand{\PO}{\ensuremath{P(O)}}
\newcommand{\POx}{\ensuremath{P(O|\mathbf{x})}}

\newcommand{\Andersen}{B.C.\,Andersen et al.\ (in prep., 2026)}

\title{Updating the PATH framework with FRB host galaxy models} % working title

\author{C.~W.~James}
\affiliation{International Centre for Radio Astronomy Research, Curtin University, Bentley, 6102, WA, Australia}
\email[Clancy W.\ James]{clancy.james@curtin.edu.au}
\author{N.~Loudas}
\affiliation{Department of Astrophysical Sciences, Peyton Hall, Princeton University, 4 Ivy Lane, Princeton, NJ 08544, USA}

\author{M.~Woodland}
\affiliation{Department of Astronomy and Astrophysics, University of California, Santa Cruz, CA 95064, USA}

\author{B.~C.~Andersen}
\affiliation{Department of Astronomy and Astrophysics, University of California, Santa Cruz, CA 95064, USA}

\author{J.~X.~Prochaska}
\affiliation{Department of Astronomy and Astrophysics, University of California, Santa Cruz, CA 95064, USA}
\alsoaffiliation{Kavli Institute for the Physics and Mathematics of the Universe, 5-1-5 Kashiwanoha, Kashiwa 277-8583, Japan}
\alsoaffiliation{Division of Science, National Astronomical Observatory of Japan, 2-21-1 Osawa, Mitaka, Tokyo 181-8588, Japan}

\author{J.L.~Hoffmann}
\affiliation{International Centre for Radio Astronomy Research, Curtin University, Bentley, WA 6102, Australia}

\author{L.~Marnoch}
\affiliation{School of Mathematical and Physical Sciences, Macquarie University, NSW 2109, Australia}
\alsoaffiliation{Astrophysics and Space Technologies Research Centre, Macquarie University, Sydney, NSW 2109, Australia}
\alsoaffiliation{Australia Telescope National Facility, CSIRO Space \& Astronomy, Box 76 Epping, NSW 1710, Australia}

\author{S.~D.~Ryder} %0000-0003-4501-8100
\affiliation{School of Mathematical and Physical Sciences, Macquarie University, NSW 2109, Australia}
\alsoaffiliation{Astrophysics and Space Technologies Research Centre, Macquarie University, Sydney, NSW 2109, Australia}


\begin{document}

\begin{abstract}
Over a hundred fast radio burst (FRB) host galaxies have now been identified, enabling both comparisons of host redshift with FRB dispersion measure to study the cosmological distribution of ionised gas, and analyses of host properties in order to identify FRB progenitors. The standard method for determining the most likely FRB host galaxy in an optical image is the Bayesian framework Probabilistic Association of Transients to their Hosts (PATH), which accounts for uncertainties in the radio localisation, and simplified prior distributions on the host being observable. In this work we extend PATH, incorporating physically-motivated priors that are based on expectations about FRB host galaxy magnitudes. We develop three different models for the apparent r-band magnitude distribution based on an FRB’s expected host galaxy redshift, $P(m_r|z)$ and combine these with expectations for redshift based on an FRB's dispersion measure, $P(z|{\rm DM})$.
We fit the parameters of these prior models using host galaxy candidates for 32 FRBs detected by the Australian SKA Pathfinder (ASKAP) in incoherent sum (ICS) mode by the Commensal Real-time ASKAP Fast Transients (CRAFT) survey.

Employing PATH with the new priors on the host magnitudes, we find increased confidence in the most probable hosts of all ASKAP ICS FRB host galaxies. All three models predict similar distributions of FRB host magnitudes at low redshift $(z \sim 0.1)$, and we confirm previous results that the true FRB host galaxy distribution is fainter than expected for a star-formation-weighted distribution (p-value of 0.12\%). However, a mass-weighted distribution provides an even worse fit (p-value of $10^{-9}$). Tests against more FRBs in the $z > 0.5$ range, where the models differ, and extensions of the models to account for e.g.\ host metallicity, may help to resolve these uncertainties in the FRB host distribution.
\end{abstract}

\maketitle

\section{Introduction}
\label{sec:intro}

Fast Radio Bursts (FRBs) are extremely energetic radio transients that reach the Earth from cosmological distances \citep{lorimer_bright_2007}. Their most likely origin is in highly magnetised neutron stars, due to the extreme energy of their bursts \citep{Ryder2023}, micro-second time-structures \citep{Farah2018}, extreme magneto-ionic environments \citep{michilli_extreme_2018}, and the bright radio flares observed from the Galactic magnetar SGR~1935+2154 \citep{2020Natur.587...59B,chimefrb_collaboration_bright_2020}. However, their local environments range from extremely bright persistent radio sources located in dwarf, star-forming galaxies \citep{2017Natur.541...58C,niu_repeating_2022}, to globular clusters, which have long since halted any star formation \citep{2022Natur.602..585K}, although this may be biased towards the repeating subset of the population, which allow VLBI follow-up. This suggests multiple pathways capable of producing FRB progenitors, each of which favours different astrophysical environments, and hence different classes of FRB host galaxies. Therefore, the cosmological evolution of the FRB population (and hence their host galaxies) may be complex, with some FRBs tracing star formation with a cosmologically insignificant delay of tens of millions of years, and potentially favouring low-metallicity environments; while others may result from long-duration mergers, and may better correlate with the total stellar mass of their host. Untangling such complexity is important both for studies of FRB host galaxies, and those that fit the redshift-dispersion measure distribution of the FRB population. In this work, we focus on the former, but note that these results are critical for the latter.

Several works have analysed FRB host galaxies and compared their properties to those of field galaxies, with a picture gradually emerging of the hosts of apparently once-off FRBs approximately tracing the star-forming main sequence \citep{2023ApJ...954...80G},  sharing similar properties to the hosts of core-collapse supernovae (CCSNe) and short-duration gamma-ray bursts (sGRBs) \citep{Bhandari+22}. However, it has been observed that there is a deficit of low-mass dwarf hosts with high star-formation rates (SFR) \citep{Bhandari2020,2020ApJ...903..152H}, especially when compared to the host galaxies of long-duration gamma-ray bursts (LGRBs) and superluminous supernovae (SLSNe) \citep{Bhandari+22}, which has been attributed to the FRB progenitors favouring higher metallicity environments \citep{DSA_Sharma_sfr}, although this is debated \citep{Loudas25,2026ApJ...997L...6Y}. \citet{2026ApJ...996...78H} argue that FRB hosts appear to trace a linear combination of stellar mass and star-formation rate (those authors also consider, and rule out, globular clusters); however, as noted by \citet{Loudas25}, there appears to be an excess of dwarf FRB host galaxies with relatively low SFR, which cannot be explained by either model. Furthermore, the sub-class of highly active FRBs associated with persistent radio sources (PRS) do appear to come from dwarf, star-forming hosts \citep[e.g.,][]{2026ApJ...996L..16M}, with PRS properties that may be evolving on year-long timescales, suggesting an origin in a young supernova remnant \citep{2026SciBu..71...76N}, and giving further evidence to a correlation with massive star formation.

All the above results rely on confident associations of FRBs with their (nominal) host galaxies. The standard methodology has been the implementation of the Probabilistic Association of Transients to their Hosts \citep[PATH;][]{PATH}, a Bayesian framework that associates candidate host galaxies identified in optical images with FRB localisations from radio telescopes. PATH integrates priors on the FRB host galaxy magnitude and offset distribution to assign posterior probabilities of each galaxy being the true host, or the true host being unseen in the image. PATH has been applied to determine the hosts of FRBs detected with the Australian SKA Pathfinder \citep[ASKAP; ][]{Shannon_ICS}, the MeerKAT telescope \citep{2026MNRAS.545f2144P}, the Deep Synoptic Array \citep[DSA; ][]{DSA_Sharma_sfr}, and the Canadian Hydrogen Intensity Mapping Experiment \citep[CHIME; ][]{2025ApJS..280....6C}.

The current application of PATH to the FRB literature has suffered two main critiques. The first is that, like any Bayesian framework, the posterior probabilities assigned by PATH are a function of the assumed prior distributions, and these were developed at a time when little was known about FRB hosts. However, this is no longer the case.
In particular, the relationship between FRB dispersion measure (DM) and host galaxy redshift ($z$), known as the `Macquart relation' \citep{Macquart2020}, is now well-studied. Several works modelling the FRB $z$--DM relation have been developed \citep{FRBPOPPY,james_zdm_2022,2023ApJ...944..105S}, all of which can model the probability distribution of an FRB's redshift from its DM, $P(z|{\rm DM})$. In particular, an FRB's DM places a stringent upper limit on its distance, since while over-dense structures in the Universe (e.g.\ galaxies, galaxy groups, halos, and filaments) can readily produce an excess of DM, nothing produces a negative DM, and even voids have a certain minimum density \citep{james_measurement_2023,2026arXiv260501994S}. In theory therefore, and almost uniquely among different classes of astrophysical transients, an FRB's DM can be used to inform prior probabilities on its host distance, and hence apparent magnitude. A consequence of this is that a low-DM FRB is much more likely to have a detectable host galaxy than a high-DM FRB, and this should be reflected in the prior probability of an FRB's true host galaxy being visible in an optical image. Similarly, the probability of seeing the true host should reflect the depth of the optical image being used to search for it.
For example, one might imagine that the prior on the true host galaxy being unseen in an image should be much lower for an FRB with a DM of 100\,\pccc\ studied with the Hubble Space Telescope than it should be for a 3500\,\pccc\ FRB searched-for in ground-based archival data. 

Quantifying the effects of optical follow-up depth, and the predictive power of an FRB's DM, requires a model for FRB host galaxies that can turn a redshift prediction, $P(z|{\rm DM})$, into a prediction for observable galaxy properties---for instance, via a prediction for r-band magnitude, $P(m_r|z)$. While such a model might usefully predict many other galaxy properties (star-formation history, metallicity, etc.), we focus on magnitude only. This is most relevant because PATH is typically used to select a likely host from a single photometric observation, with more-detailed optical follow-up observations subsequently yielding further data. Despite the many aforementioned works analysing the specific properties of FRB hosts however, relatively few \citep[e.g.][]{Marnoch2023,Loudas25,2026ApJ...996...78H} have developed specific models of FRB hosts with which to compare to data. Ideally, developing such a model would not only enable the use of an FRB's DM to create informed priors for host identification with PATH, it would also allow model parameters to be fit to the subsequent FRB host distribution, providing information on the progenitors of FRBs.

The second critique relates to the interpretation of PATH output, in particular the tendency to separate host galaxy candidates into `firm' hosts if their posterior probability of being the true host is greater than e.g.\ 90\%, and to discard hosts lower than this threshold. However, any Bayesian framework can only produce an unbiased posterior distribution when summing over all outcomes---in this case, candidate host galaxy properties---weighted by their posterior probabilities. This has recently been demonstrated in the case of PATH by \Andersen. While we cannot correct for this bias in this work, since doing so would require spectroscopic observations of a large number of FRB host candidates, we consider it important to estimate the degree of bias in the current literature produced by this approximation.

We first therefore review and expand the PATH framework in \S\,\ref{sec:path}, before developing models of the FRB host galaxy apparent magnitude distribution as a function of redshift $P(m_r|z)$ in \S\,\ref{sec:models}. In \S\,\ref{sec:fitting}, we combine these with estimates of FRB redshift, $P(z|{\rm DM})$, and fit these models to optical follow-up observations of 32 FRBs observed with ASKAP in incoherent sum (ICS) mode by CRAFT. We present our fit results, together with updated estimates of the intrinsic FRB host galaxy distribution, and probabilities for galaxies being the true host, in \S\,\ref{sec:results}. Finally, in \S\,\ref{sec:discussion}, we discuss the implications of our results.

\section{PATH, revisited}
\label{sec:path}

In this Section, we briefly recap the standard PATH framework (a more comprehensive review can be found in \Andersen), before presenting our updates to the methodology. We first show how priors on host galaxy magnitudes can be incorporated into the standard PATH framework, before revisiting the calculation of PATH posteriors, and concluding with a final note on the correct normalisation of candidate priors. In this Section, we use a generic $m$ to indicate an apparent magnitude in an arbitrary band; later, we will specifically use $r$-band magnitudes $m_r$. However, we note that, by replacing $m$, the below analysis could readily be applied to any other host property. Throughout, we refer readers to Table~\ref{tab:notation}, to track both original PATH notation, and new notation introduced here.

\begin{table*}[]
    \centering
    \begin{tabular}{ccl}
    \multicolumn{3}{l}{Original PATH terminology} \\
    \hline
    Symbol & Unit & Description \\
    \hline
     $m$ & mag & Apparent magnitude \\
     $m_r$ & mag & Apparent r-band magnitude \\
     $\Sigma m_i$ & Gal arcsec$^{-2}$ & Cumulative density of galaxies with magnitude $m < m_i$ on the sky \\
     $x$ & deg & Radio localisation position (RA,DEC) of an FRB on the sky \\
     $\phi$ & arcsec & Galaxy half-light radius \\
     $\theta$ & arcsec & Observed offset between FRB and host candidate centre \\
     $\omega$ & arcsec & True offset between FRB and its host \\
      $O_i$ && The $i^{\rm th}$ host galaxy candidate \\
      $P(O_i)$   &  & Prior of host galaxy candidate $i$ being the true host \\
      $P(x|O_i)$ & & Probability of observing FRB position $\vec{x}$ given that host candidate $O_i$ is the true host \\
      $P(O_i|x)$   & & Posterior probability of host galaxy candidate $i$ being the true host, given FRB localisation $x$ \\
      $P(U)$   & & Prior of the true host galaxy being unidentified in an image \\
      $P(U|x)$  & & Posterior probability of the true host galaxy candidate being unidentified in an image, given FRB localisation $x$ \\
      \hline
      \hline
    \multicolumn{3}{l}{Updated terminology from this work} \\
    \hline
    Symbol & Unit & Description \\
    \hline
      $\rho_{D16}(m)$ & Gal arcsec$^{-2}$ mag$^{-1}$& Intrinsic angular density of galaxies with magnitude $m$ on the sky, taken from \citet{Driver2016} \\
      $\rho(m)$ & Gal arcsec$^{-2}$ mag$^{-1}$& Apparent angular density of galaxies with magnitude $m$ on the sky, as identified by an optical system \\
      $\Omega$ & arcsec$^2$ & Angular size of optical image about FRB position \\
      $N_O$ & & Number of host galaxy candidates \\
      $\mathbf{x}$ &  & Set of positions of all FRB host galaxy candidates in an image, relative to the FRB localisation $\vec{x}$.\\
      $P(\mathbf{x},N_O)$& & Probability of observing a set of $N_O$ host 
      galaxy candidates with positions $\mathbf{x}$ \\
      $P(\mathbf{x},N_O|O_i)$ & & Probability of observing a set of $N_O$ host galaxy candidates with positions $\mathbf{x}$ given that $O_i$ is the true host \\
      $P(O_i|\mathbf{x},N_O)$ && Posterior probability of the true host being galaxy $O_i$ given $N_O$ host galaxy candidates with positions $\mathbf{x}$ \\
      $P(\mathbf{x},N_O|U)$ & & Probability of observing a set of $N_O$ host galaxy candidates with positions $\mathbf{x}$ given that the true host is unseen \\
      $P(U|\mathbf{x},N_O)$ && Posterior probability of the true host being unseen given $N_O$ host galaxy candidates with positions $\mathbf{x}$ 
    \end{tabular}
    \caption{PATH notation used in this work. The upper portion summarises notation from the original PATH work of \citet{PATH}; the lower portion, new notation introduced here.}
    \label{tab:notation}
\end{table*}

\subsection{Recap of the original PATH framework}
\label{sec:original_path}

The PATH framework uses two sets of priors. One is on the radial offset distribution with respect to the centre of the true host, $P(\omega)$. This is typically modelled as an exponential in terms of the host galaxy's angular size, $\phi$, and weighted by a factor $2 \pi \omega$ to account for the greater solid angle at larger radii. Recent data prefer a scale length of $\phi/2$ \citep{Shannon_ICS}. This work however focuses on the second prior, $P(O_i)$, giving the probability of a particular galaxy $O_i$ being the true host given its apparent magnitude, $m$. The standard host prior is the `inverse' prior,
\begin{eqnarray}
    P(O_i) & = & \frac{1}{\Sigma m_i}, \label{eq:pmr}
\end{eqnarray}
where $\Sigma m_i$ is the angular number density of galaxies with magnitude $m \le m_i$ on the sky, taken from \citet{Driver2016}. A prior on the true host being unobserved, $P(U)$, is defined, and the priors $P(O_i)$ get re-normalised to ensure that
\begin{eqnarray}
    \sum_i P(O_i) & = & 1 - P(U). \label{eq:orig_path}
\end{eqnarray}
Given these priors, PATH calculates Bayesian posteriors according to 
\begin{eqnarray}
    P(O_i|x) & = & \frac{P(x|O_i) P(O_i)}{P(x)}, \label{eq:path_posteriors1}\\
    P(U|x) & = & \frac{P(x|U) P(U)}{P(x)}, \label{eq:path_posteriors2}
\end{eqnarray}
where $x$ represents the observable information, i.e.\ the FRB localisation relative to galaxies identified in the optical image. The term $P(x|O_i)$ accounts for the offset prior $P(\omega)$, the FRB localisation ellipse, and the position and size of host candidate $i$ with respect to that localisation. As we do not address this factor in this work, we refer readers to \citet{PATH} for further details of its calculation. The factor $P(x|U)$ is simply set to $\Omega^{-1}$, i.e.\ the inverse size of the optical image, $\Omega$. Thus the normalising factor $P(x)$ is calculated as
\begin{eqnarray}
P(x) & = & P(x|U) P(U) + \sum_i P(x|O_i) P(O_i). \label{eq:px}
\end{eqnarray}

\subsection{Re-evaluating the probability of an observation within PATH}
\label{sec:likelihood}

Here, we derive a new formula for the probability of an observation, which affects how PATH posteriors are calculated. We ignore galaxy clustering in our derivation, and treat all galaxy observations as independent. We comment on this assumption below. We first note that the `observation', in a Bayesian sense, includes both the radio localisation information from the FRB detection, relative to the locations of candidate galaxies, and the existence of $N_O$ host candidates in the analyzed image area $\Omega$. Here, we explicitly separate these terms, and instead of using the FRB localisation $x$, we define $\mathbf{x}$ to be the set of positions of host candidates relative to the FRB localisation $x$, while $N_O$ refers to the number of candidates, for reasons that will become apparent later. All probabilities in this Section are functions of candidate host magnitudes, so we do not explicitly write this dependence in the equations. The usual normalisation applies, being that the probability of this observation is equal to the sum of the probabilities that each individual galaxy is the true host, plus the probability that none of them are the true host, i.e.\
\begin{eqnarray}
    P(\mathbf{x},N_O) & = & \sum_i P(\mathbf{x},N_O|O_i) P(O_i) + P(\mathbf{x},N_O|U) P(U), \quad\label{eq:likelihood1}
\end{eqnarray}
where $P(\mathbf{x},N_O|O_i)$ is the probability of the observation given that galaxy $O_i$ is the true host, and $P(\mathbf{x},N_O|U)$ is the probability of that observation given that the true host is unobserved. These disjoint scenarios are weighted against the usual priors, $P(O_i)$ and $P(U)$.

The term $P(\mathbf{x},N_O|O_i)$ includes the standard term $P(\mathbf{x}|O_i)$ from PATH; however, if $O_i$ is the true host, the presence of all other galaxies must be coincidental (in the regime where galaxy clustering is ignored). Thus we find
\begin{eqnarray}
   P(\mathbf{x},N_O|O_i) & = & P(\mathbf{x}|O_i) \,\Pi_{j=1,j \ne i}^{N_O} \rho(m_j), \label{eq:lxoi}
\end{eqnarray}
where $\rho(m_j)$ is the angular number density on the sky of all galaxies having magnitude $m_j$, i.e., it is the derivative of $\Sigma m_j$. Since all $j \ne i$ galaxies are not the true host, their positions in the image with respect to $\mathbf{x}$ hold no information, and thus have a uniform probability density of $P(\mathbf{x}) = \Omega^{-1}$, where $\Omega$ is the analysed image area; however, the probability of that galaxy being present in the image is $\Omega \rho(m_j)$, so that these factors of $\Omega$ cancel. Given that $O_i$ is the host, its presence is guaranteed, so the probability of it being in the image is unity.

Similarly, given that the true host is unseen, the probability of seeing that configuration of galaxies is
\begin{eqnarray}
      P(\mathbf{x},N_O|U) & = &  \Pi_{j=1}^{N_O} \rho(m_j), \label{eq:lu}
\end{eqnarray}
since the presence of each must be coincidental. We can now write Equation~\ref{eq:likelihood1} as  
\begin{eqnarray}
    P(\mathbf{x},N_O) & = & \sum_i \left( P(\mathbf{x}|O_i)\, \Pi_{j=1,j \ne i}^{N_O} \rho(m_j)  \right) P(O_i) \nonumber \\
    & & + \Pi_{j=1}^{N_O} \rho(m_j) P(U) \nonumber\\
    & = & \Pi_{j=1}^{N_O} \rho(m_j) \left[  \sum_i \frac{P(\mathbf{x}|O_i) P(O_i)}{\rho(m_i)}  + P(U) \right]. \label{eq:likelihood2} 
\end{eqnarray}
Comparing Equation~\ref{eq:likelihood2} to the standard PATH framework, we observe firstly that the leading normalising constant is the probability of all galaxies being present in such an image irrespective of the FRB host considerations. It can therefore be ignored entirely in the calculation of posteriors, and for likelihood maximisation purposes, since it is a common factor in all calculations. Next, we see that the term $P(\mathbf{x}|O_i) P(O_i) \rho(m_i)^{-1}$ is almost identical to Equation~\ref{eq:path_posteriors1}, except for the replacement of $\Sigma m_i$ with its derivative, $\rho(m_i)$. Lastly, we note that the prior $P(U)$ appears in unmodified form. Effectively, this implies that $P(\textbf{x}|U)=1$, i.e.\ that these foreground galaxies were always going to exist irrespective of the FRB localisation.

Removing therefore the common leading factor from Eqs.~\ref{eq:lxoi}--\ref{eq:likelihood2} gives
\begin{eqnarray}
    P(\mathbf{x},N_O) & = & \sum_i \frac{P(\mathbf{x}|O_i) P(O_i)}{\rho(m_i)}  + P(U), \label{eq:likelihood3} \\
     P(U|\mathbf{x},N_O) & = &  \frac{P(U)}{P(\mathbf{x},N_O)}, \label{eq:puposterior} \\
     P(O_i | \mathbf{x},N_O) & = & \frac{P(\mathbf{x}|O_i) P(O_i)}{P(\mathbf{x},N_O) \rho(m_i)}. \label{eq:poposterior}
\end{eqnarray}

\subsection{A comment on PATH prior normalisation}
\label{sec:normalisation}

A key (and fairly obvious) identity should hold in all stages of a PATH analysis: that the probabilities of the host being seen, and unseen, should sum to unity. In the original PATH work, this is applied via Equation~4 of \citet{PATH}, which we have written as Equation~\ref{eq:orig_path}. However, writing this equation in this manner misses one key subtlety which only becomes obvious when constructing priors based on host galaxy magnitudes. This is that priors should be set before looking at an image; once a list of candidate hosts $O_i$ have been identified in an image, any priors should be updated to reflect that observation (i.e., they are not `priors' anymore)! This becomes critical once expectations for FRB host galaxy properties are set: if one expects FRB hosts to be bright and star-forming, but only a red-and-dead dwarf galaxy is observed in an image, then a reasonable observer would increase their expectation of the true host being unobserved. However, within the original PATH framework, the sum of priors on all observed galaxies gets renormalised to $1-P(U)$ via Equation~\ref{eq:orig_path}. Thus an optical image with a single unlikely host candidate would have that candidate's prior renormalised to $1-P(U)$, as would an optical image with a single likely host candidate.

As it turns out, this distinction did not matter for the original PATH analysis, since the priors $P(U)$ were by construction rather generic, and not based on any particular expectation of the nature of the FRB host galaxy. In that sense, this behaviour was a feature, rather than a bug, and ensured that the DM--$z$ relation found by \citet{Macquart2020} was not influenced by such priors. Indeed, a generic prior of e.g.\ $P(U)=10$\% could not be updated once a list of candidates $O_i$ was found in an image, because there was no framework to determine if, having observed those candidates, $P(U)$ should become more or less likely. However, for this work, this distinction now becomes critical.

Here, we construct priors $P(m)$, which is the probability that the true host has apparent magnitude $m$. We ensure that our prior expectations are correctly normalised by requiring that
\begin{eqnarray}
    \int P(m) dm & = & 1.
\end{eqnarray}
We also construct a function $P(O|m)$ in \S\,\ref{sec:pogm}, such that
\begin{eqnarray}
    P(O) & = & \int  P(O|m)  P(m) d m \label{eq:PO}\\
    P(U) & = & \int (1-P(O|m)) P(m) d m, \label{eq:PU}
\end{eqnarray}
which ensures correct normalisations on the priors that the true host is or is not seen in an image. The prior $P(O_i)$ that a particular galaxy of magnitude $m_i$ is the host is therefore the integrand of Equation~\ref{eq:PO}, i.e.\
\begin{eqnarray}
    P(O_i) & = & P(O|m_i)  P(m_i).
\end{eqnarray}
This effect must also be accounted-for in the galaxy density function, $\rho(m_i)$. Thus we use
\begin{eqnarray}
    \rho(m_i) & = & P(O|m_i) \rho_{\rm D16}(m_i)
\end{eqnarray}
where $\rho_{\rm D16}(m_i)$ is the galaxy density function from \citet{Driver2016}. In effect, this means that factors $P(O|m_i)$ in both the numerator and denominator of Equation~\ref{eq:likelihood3} cancel, and this term affects the likelihood only through $P(U)$---which then indirectly affects $P(O_i|\mathbf{x},N_O)$ in Equation~\ref{eq:poposterior}.

Out of curiosity, we derive an updated version of Equation~\ref{eq:orig_path}, i.e., a set of updated priors once FRB host galaxies are seen in an image, but before positional information is included. This is accomplished by removing all positional information from Equation~\ref{eq:likelihood2}, obtaining
\begin{eqnarray}
    P(N_O) & = & \Pi_{j=1}^{N_O} \Omega \rho(m_j) \left[ \sum_i \frac{P(O_i)}{\Omega \rho(m_i)}  + P(U) \right], \label{eq:pno}\\
    P(O_i|N_O) & = &  \frac{1}{P(N_O)} \sum_i \frac{P(O_i)}{\Omega \rho(m_i)}, \label{eq:pono}\\
    P(U|N_O) & = & \frac{P(U)}{P(N_O)}.
    \label{eq:puno}
\end{eqnarray}
where the factors $\Omega$ have re-entered the equation, since they are no longer cancelled by $P(\mathbf{x}) = \Omega^{-1}$ for galaxies which are not the host. In an extreme case, where the $\Omega$ is extremely small, the presence of even a single galaxy will make that the likely host. Conversely, when $\Omega$ is extremely large, the summation over individual $O_i$ becomes an integral over the observable galaxy magnitude distribution $P(m)$, with a density equal to the expected background density
\begin{eqnarray}
\sum_i \frac{P(O_i)}{\Omega \rho(m_i)} & \Rightarrow & \frac{1}{\Omega} \int \frac{n(m) P(O|m) P(m)}{\rho(m)} dm \nonumber \\
& \sim & \frac{1}{\Omega} \int \frac{\Omega {\rho(m)}  P(O|m) P(m)}{\rho( m)} dm  \\
& = & \int P(O|m) P(m) dm \nonumber \\
& = & P(O),
\end{eqnarray}
and thus the priors will not be updated, as might be expected.

\subsection{Qualitative effects of clustering}

We finally comment on the qualitative effects of galaxy clustering. Firstly, our priors $P(U)$ and $P(O_i)$ will be unaffected by clustering, since whether or not the true host is visible in an image is determined only by the properties of the host. It will however affect the calculation of posterior probabilities. Yet, the effect may be small, as can be demonstrated with this trivial example. Consider a Universe with all galaxies having equal magnitude, and a galaxy density of one per ten square arcseconds ($\rho = 0.1$). However, galaxies are clustered such that one in every hundred square arcseconds has ten galaxies, i.e.\ $\rho_{\rm cluster}=0.01$. Observing a cluster of $N_O=10$ galaxies in an optical image of $1^{\prime\prime}$ about an FRB localisation should therefore increase the posterior of observing the true host, $P(O|N_O)$, by the inverse of the cluster density, $\rho_{\rm cluster}^{-1}=100$, relative to $P(U|N_O)$, since if the true host (and hence, the host cluster) is unobserved, there is a 1\% chance of coincidentally observing a cluster along that line of sight. If instead we ignore clustering and use Equation~\ref{eq:pno}, each individual galaxy in the cluster will have the posterior updated by a factor of $\rho^{-1} = 10$. Yet this will be summed over all ten cluster galaxies, yielding the correct factor of $100$. Whether or not such equality holds for non-trivial examples remains an open question. However, this does suggest that ignoring clustering will not significantly skew our posterior calculations.

\section{Host galaxy models}
\label{sec:models}

Galaxies are complex. Within the context of FRB research, their primary observables are their apparent magnitude and angular extent, by which they are assessed as host galaxy candidates by PATH as discussed above; and their redshift, which allows them to be placed on the z--DM relation. However, a whole host of properties can be extracted from spectroscopic observations, such as stellar mass, star-formation rate (SFR), metallicity, and star-formation history \citep[e.g.\ ][]{2023ApJ...954...80G}, while radio observations inform us of HI and continuum emission \citep{2024ApJ...962L..13G}.

In this initial work, we aim to model only a function $P(m_r|z)$, giving the distribution of apparent $r$-band magnitudes $m_r$ as a function of redshift. This is because most efforts to identify FRB hosts rely only on a single observation, typically with an $r$-band filter (here, we approximate $R$-band and $r$-band filters as being identical, and use `$r$' to refer to both). We develop three different models, and then compare their predictions against observations of FRBs with ASKAP. %Example outputs of these models are given in Figure~\ref{fig:pmrgz}.

\subsection{Model 1: Empirical (`Marnoch23')}
\label{sec:marnoch}

Our empirical model (which we refer to as the ``Marnoch23'' model) is that presented by \citet{2026arXiv260300371J} based on the work of \citet{Marnoch2023}, who derive the $m_r(z)$ curves for 23 FRB host galaxies. The model uses a Gaussian, with redshift-dependent parameters, to characterise the $m_r(z)$ distribution, with mean $\mu(z)$ and standard deviation $\sigma(z)$ fit to the mean and RMS of the curves derived by \citet{Marnoch2023}. The key advantage of this model is that no assumptions about the FRB host galaxy distribution need to be made. Rather, it just uses the available information for actual FRB hosts, accounting for all their complexity, irrespective of whether or not we as astronomers understand it. It also naturally includes $K_r-$corrections for each host individually, since the spectra were fit with {\sc PROSPECTOR} \citep{2017ApJ...837..170L}. However, the model does not account for the evolution of galaxies: FRB hosts may look quite different at $z=1$ to those at $z=0$, and this sample all had hosts within $z<0.5$. Furthermore, while  17 of the FRB hosts used in the model were localised by the CRAFT collaboration, and represented a mostly complete sample (with the exception of the FRB in question, FRB~20210912A, which at time of publication did not have any known host \citep{Marnoch2023}) it also included six others of mixed provenance, e.g.\ FRB~20190520B, which was localised only because it is a strong repeater \citep{niu_repeating_2022}. Finally, the model is based on `firm' hosts, and thus is subject to all the biases that we discuss in \S\,\ref{sec:intro}---indeed, it was developed precisely because the host of FRB\,20210912A was too faint to be visible in optical and near-infrared follow-up observations. We also note that the model is completely specified: there are no free parameters which would allow the model to be adjusted or fit to further data, e.g.\ to hosts too faint to be included in the original sample of 23 FRBs.

\subsection{Model 2: Predictive (`Loudas25')}
\label{sec:loudas}

\begin{figure}
    \centering
    \includegraphics[width=\linewidth]{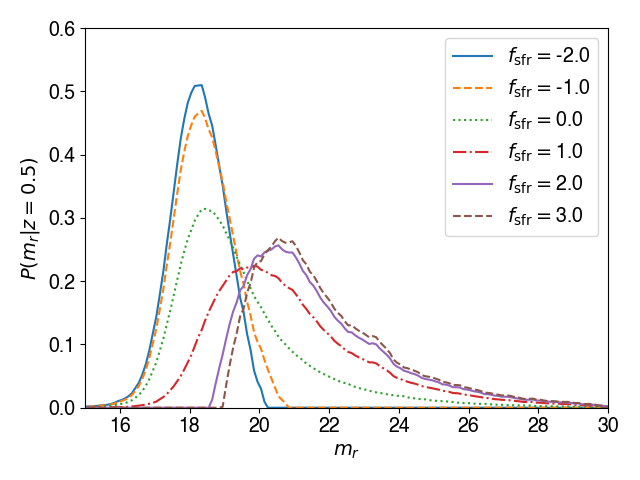}
    \caption{Probability distributions $P(m_r|z)$ at $z=0.5$ for various values of $f_{\rm sfr}$ using the Loudas25 model.}
    \label{fig:loudas_example}
\end{figure}

Our second model is based on {\sc GALFRB}\footnote{\url{github.com/loudasnick/GALFRB}}, a package developed by \citet{Loudas25} to generate galaxy populations that trace various physical quantities, such as $\rm SFR$ and/or $\rm M^*$, while accounting for redshift evolution. {\sc GALFRB} is built on the probability density function $\rho({\rm SFR},M^*, z)$, given by the trained neural network (NN) presented by \citet{2022ApJ...936..165L}, and is based on data from the COSMOS-2015 \citep{2016ApJS..224...24L} and 3D-HST \citep{2014ApJS..214...24S} UV-IR catalogues in the redshift range 0.2--3. {\sc GALFRB} incorporates analytical expressions to complete the model in the low-$z$, low-$\rm M^*$ range, where data are insufficient to train the NN. The $r$-band magnitudes are sampled for each galaxy by modeling the full rest-frame mass-to-light ratio distribution via the $g-r$ colour which in turn depends on SFR; these values are then $K_r-$corrected according to \citet{2023ApJ...951..137K}, yielding apparent magnitude values. We refer readers to \citet{Loudas25}, and references contained therein, for further details of the modelling.

We generate redshift-dependent magnitude distributions, $P(m_r|z)$, by sampling mock galaxies in a redshift grid with resolution $\Delta z = 0.01$ from $0.01$-$2$.  Selecting galaxies weighted by their stellar mass $M^*$ produces $P_{\rm M^*}(m_r|z)$, while weighting by their star-formation rate produces $P_{\rm SFR}(m_r|z)$. We smoothly scale between these scenarios by using the factor $f_{\rm sfr}$, producing
\begin{eqnarray}
P(m_r|z) & = & f_{\rm sfr} P_{\rm SFR}(m_r|z)  + (1-f_{\rm sfr}) P_{\rm M^*}(m_r|z).
\end{eqnarray}
Probabilities at redshifts less than $z=0.01$, and greater than $z=2$, were extrapolated by scaling and shifting the distribution corresponding to the lowest/highest redshift bin according to the relative luminosity distance only. Example distributions of $P(m_r|z)$ are given in Figure~\ref{fig:loudas_example}. The $M^*$-weighted distribution ($f_{\rm sfr}=0.0$) predicts brighter hosts than the SFR weighted distribution ($f_{\rm sfr}=1.0$); thus, higher $f_{sfr}$ values yield fainter galaxy populations. Extrapolating the distributions to very low ($f_{\rm sfr} < -2$) or very high ($f_{\rm sfr} > 3$) values produces little change in the predictions, so we only consider the range $-2 \le f_{\rm sfr} \le 3$ in this work.

This model has several advantages. Firstly, it is based on a large sample of galaxies, and thus accounts for redshift evolution of the galaxy population. Its direct dependence on SFR and $M^*$ allows us to test predictions for FRB host galaxies in a manner that informs us about FRB origins. Consistency with $f_{\rm sfr}=1$ would clearly indicate that FRBs trace star-formation, and may originate in CCSNe; while $f_{\rm sfr}=0$ would perhaps indicate a merger origin with a very large lag compared to star formation. An intermediate value may indicate a mixed population. Even though $\rm f_{sfr}$ has a physical meaning only in the range $\rm 0\leq f_{sfr} \leq 1$, we relaxed this condition by numerically allowing for unphysical values of $f_{\rm sfr}>1$ and $f_{\rm sfr}<0$, by setting $P(m_r|z)$ to zero if it would otherwise become negative, although the interpretation of such values is unclear. See \S\,\ref{sec:discussion} for further discussion of such scenarios, in light of our results.

A potential disadvantage of the model is that FRB hosts can only be sampled according to $M^*$ and SFR, so that effects such as metallicity, moderate time-delays with respect to star formation, or other galaxy properties that may influence FRB creation cannot be explicitly accounted for. We note that metallicity could in principle be incorporated, as it is a function of SFR and $M^\star$ \citep{2026ApJ...997L...6Y}, but we leave this to a future work.

\subsection{Model 3: Naive (this work)}
\label{sec:naive}

Our final model is termed the ``Naive'' model, because it is not based on any astronomical observations. Rather, it models FRB hosts as having a fixed distribution in absolute magnitude $M_r$, which is then scaled to apparent magnitude $m_r$ using only the luminosity distance, and a K-correction-like term. We parameterise the $M_r$ distribution using six values between $M_r = -25$ and $M_r = -15$, and use linear interpolation to infer the full $P(M_r)$ distribution (see \ref{sec:binning} for a justification of this choice). We include a simple redshift dependence which acts both as a $K_r$-correction and evolutionary correction, so that
\begin{eqnarray}
m_r & = & M_r + 5 \log_{10} L_d -5-K_r
\end{eqnarray}
where the redshift dependence is encapsulated in the nominal K-correction
\begin{eqnarray}
    K_r & = & 2.5 k \log_{10} (1+z).
\end{eqnarray}
In reality, the redshift dependence will be much more complicated: galaxy evolution will affect the distribution $P(M_r)$ through various processes, introducing a complex $z$-dependence, and the actual $K_r$-correction will also be more complex. For now, we instead introduce only a single term, and test for the significance of $k \ne 0$ in our fits.

The advantage of this model is that it is generic---there are no prior assumptions about FRB astrophysics, and it therefore cannot be influenced by any of the biases discussed in this work. The disadvantages of the model are obvious: it can only account for galaxy evolution and spectra via a single $K_r$-correction term applied to the model as a whole, which vastly over-simplifies the complexities of the galaxy population. Furthermore, it will be difficult to relate model results to astrophysical conclusions about FRB origins.

\section{Evaluation}
\label{sec:fitting}

\subsection{Dataset, and $P(O|m)$}
\label{sec:pogm}

We use optical follow-up images for 32 of the ASKAP/CRAFT incoherent sum (ICS) survey FRBs \citep{Shannon_ICS}, which were centred on the FRB radio localisations. This includes all available data from this sample, with the exceptions of FRB20230731A and FRB20230718A, which lie within the Galactic plane. The images were taken with the FOcal Reducer and low dispersion Spectrograph \citep[FORS2; ][]{FORS2} on the ESO Very Large Telescope (VLT), and processed as described in \citet{Marnoch2023}. Approximately half the targets were observed in the $R$-band, with the remainder in $g$ and $I$ band. We convert these latter data to the $R$-band as described in \S\,\ref{sec:color}.

\begin{figure}
    \centering
    \includegraphics[width=\linewidth]{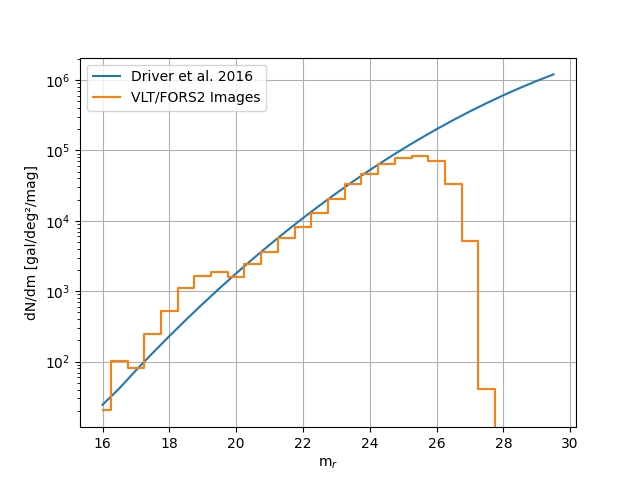}
    \caption{Galaxy number density per magnitude bin identified in images about the nominal positions of CRAFT/ICS FRBs. Also shown are the galaxy densities from \citet{Driver2016}.}
    \label{fig:maghist}
\end{figure}

Galaxies are identified in each optical image using {\sc SExtractor} \citep{2020ASPC..527..461B,2020ASPC..527...29K}, with {\sc DE}-{\sc TECT\_MINAREA}$ = 5$ and {\sc DETECT\_THRESH}$ = 1.5$. 
Only galaxies identified within a $1^\prime$ cutout of the FRB source are considered as plausible hosts (the largest localization uncertainty for the sample is $\approx 2''$ for FRB\,20190711A), and kept for further analysis.

Figure~\ref{fig:maghist} shows the distribution of optical magnitudes of all FRB host candidates in these images, plotted against the $r$-band density function from \citet{Driver2016}. It is evident that, compared to expectations from field galaxies, the observations become incomplete beyond magnitude $m_r=25$, with the dimmest identifiable galaxy being at $m_r=27$. The excess of galaxies in the $18 \le m_r \le 20$ range is due to mis-classified stars. We remove these by hand if they occur in the vicinity of the FRB, but we have not done this over the entirety of all images. They thus represent a small contamination to Figure~\ref{fig:maghist}, which we estimate to be negligible for purposes of estimating completeness.

\begin{table}[]
    \centering
    \begin{tabular}{r |c c}
    Optical data & $\mu_m$ & $\sigma_m$ \\
    \hline
      VLT/FORS2   & 26.2 & 0.34  \\
      Legacy Surveys   &  24 & 0.55   \\
      Pan-STARRS    & 21.8  & 0.54  \\
    \end{tabular}
    \caption{Fits of Equation~\ref{eq:pogm} to data from \Andersen\ and FORS2 data from this work.}
    \label{tab:pogm}
\end{table}

\begin{figure}
    \centering
    \includegraphics[width=\linewidth]{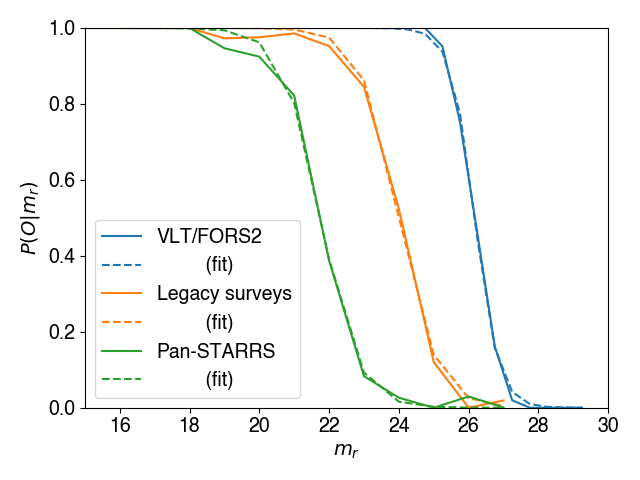}
    \caption{Data and fits to the probability $P(O|m_r)$ of a galaxy being identified in optical images given its $r$-band magnitude, $m_r$. Data for the Legacy Surveys and Pan-STARRS from \Andersen; FORS2 from this work. Fits according to Equation~\ref{eq:pogm} with parameters from Table~\ref{tab:pogm}.}
    \label{fig:pogm}
\end{figure}

We next introduce the last piece of our framework, which calculates the probability that a galaxy of magnitude $m_r$ is identified in a given optical image, i.e., $P(O|m_r)$. To estimate this, we use a procedure similar to that of \Andersen.
We use {\sc SExtractor} to extract sources from the VLT/FORS2 images, and exclude any source from our sample with {\sc CLASS\_STAR}$>0.8$. In addition, any object with a magnitude $m > 23$ is classified as a galaxy.
We then construct the observed angular density function $\rho_{\rm  VLT/FORS2}(m_r)$, and calculate the ratio between this and the expected number density $\rho(m_r)$ from \citet{Driver2016}. A small scaling factor is applied so that the completeness of each is equal at bright magnitudes.
The resulting ratio is shown in Figure~\ref{fig:pogm}, together with equivalent data from the Panoramic Survey Telescope and Rapid Response System Survey \citep[Pan-STARRS;][]{ps1} and the DESI Legacy Imaging Surveys \citep[hereafter the Legacy Surveys;][]{dlis} taken from \Andersen. We parameterise this distribution with the expression
\begin{eqnarray}
    P(O|m_r) & = & \frac{1}{1 + \exp [(\mu_m - m_r)/\sigma_m]} \label{eq:pogm}
\end{eqnarray}
which provides a good fit to both data from \Andersen\ and VLT/FORS2. The best-fit values are reported in Table~\ref{tab:pogm}.

\subsection{Colour correction}
\label{sec:color}

\begin{figure}
    \centering
    \includegraphics[width=\linewidth]{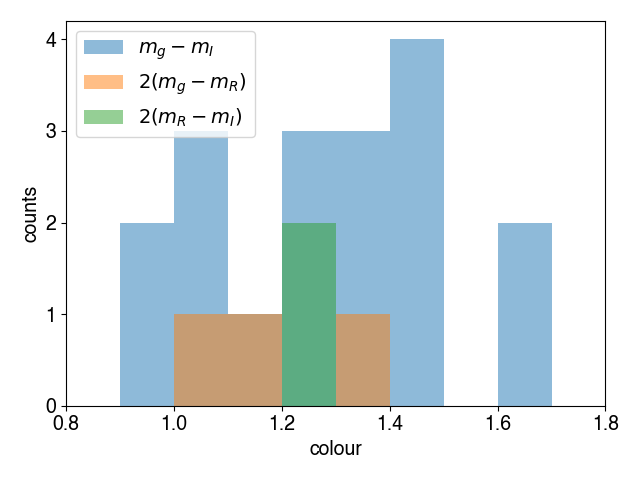}
    \caption{Histograms of VLT/FORS2 colours for ASKAP/ICS FRB hosts identified by \citet{Shannon_ICS}.}
    \label{fig:colours}
\end{figure}

The VLT/FORS2 photometry used in this work was taken predominantly in either $g$ and $I$ bands, or $R$ band. Figure \ref{fig:colours} shows $g-I$ colours for the 18 ICS FRBs with data taken in both bands, for the most likely FRB hosts identified by \citet{Shannon_ICS}. It also shows $R-I$ and $g-R$ colours for the 2 and 4 FRBs with data in both of those respective bands, multiplied by two to account for these bands being closer together than $g-I$, for visualization purposes. The mean $g-I$ colour for the sample is 1.3, with the $R-I$ and $g-R$ colours having means of 0.6 and 0.7 respectively. Galaxy surveys reveal a bimodal distribution in colour, due to the star-forming and quenched populations, with green valley galaxies in-between \citep{2003ApJ...594..186B}. FRB hosts tend to follow the star-forming main-sequence, with a small fraction transitioning to quiescence \citep{2023ApJ...954...80G}. This explains the relatively small variance in observed $m_r-m_I$ and $m_g-m_R$; however, our small sample sizes are inadequate to fully capture the variance. When running our PATH analysis therefore, we assume constant corrections of $m_g-m_R=0.65$ and $m_R-m_I=0.65$, and use these to convert galaxy magnitudes from the $g$-band and $I$-band images to their $R$-band equivalents. We estimate that this leads to errors of $\sim 0.3$ on $m_r$.

\subsection{Procedure for estimating $P(z)$}
\label{sec:zdm}

The extragalactic dispersion measure distribution of FRBs has been shown to correlate with their redshift. This was first indicated by \citet{Shannon_FE}, who showed that the brighter (and therefore, likely closer) FRBs had lower DMs, and then most famously by \citet{Macquart2020}, who proved that the DMs of FRBs increase with their redshift according to the expected baryonic density of the Universe. Many works have since attempted fits of the z--DM relationship, which (when done properly) include estimates of the detection bias against higher DMs, the FRB luminosity function and source evolution, as well as accounting for fluctuations in the density of ionised gas \citep{Connor2019,2020MNRAS.494..665L,2023ApJ...944..105S}. Nonetheless, estimates for the FRB z--DM distribution remain model-dependent, with choices such as the functional form of the luminosity function, source evolution, or frequency dependence of FRBs affecting the conclusions drawn about the population \citep{2025PASA...42...17H}. Of specific note here is that these estimates are typically based upon identified FRB host galaxies, the validity of which we explicitly set out to test in this work. In the future, we aim to simultaneously incorporate completeness in FRB hosts into FRB population and cosmological parameter fitting.

In this work, we use the \zdm\ code of \citet{james_zdm_2022} to predict the $P(z|{\rm DM}_{\rm EG})$ distribution for each FRB. The extragalactic contribution to the DM budget, ${\rm DM}_{\rm EG}$, is calculated by subtracting local contributions from the Milky Way's interstellar medium \citep[estimated using the NE2001 model of][]{cordes_ne2001} and an assumed halo contribution of 68\,pc\,cm$^{-3}$ \citep{HoffmannHalo26} from the FRB's measured DM. In the near future, we plan to update \zdm\ to the NE2025 model \citep{ne2025}. However, this has not yet been implemented within \zdm.

We assume standard Planck cosmology \citep{planck_collaboration_planck_2016}, and take FRB parameters such as the luminosity function and population evolution from \citet{HoffmannHalo26}. This is exactly the same procedure as we use in \citet{2026arXiv260300371J} to calculate predictions for Vera C.\ Rubin's Legacy Survey of Space and Time using the Marnoch23 model.

For each of three characteristic ASKAP/ICS central frequencies (900\,MHz, 1300\,MHz, and 1600\,MHz) therefore, we create best-fitting z--DM grids as per \citet{james_zdm_2022}, which produces predictions for $P(z|\DMeg)$. Our priors thus become
\begin{eqnarray}
    P(O_i) & = & \int P(O|m_i) \, P(m_i|z) \, P(z|\DMeg) dz,\\
    P(U) & = & \iint \left[ 1-P(O|m_r) \right] \, P(m_r|z) \, P(z|\DMeg) dz \,dm_r. \nonumber
\end{eqnarray}
This is calculated for each FRB individually, where we use each of the three models for $P(m_r|z)$ described in Section~\ref{sec:models}.

\section{Results}
\label{sec:results}

Our model fitting procedure aims at maximising the likelihood of an observation, $P(\mathbf{x},N_O)$, given by Equation~\ref{eq:likelihood3}. We evaluate the likelihood on a sample of 32 FRBs with optical data detected during CRAFT ICS observations \citep{Shannon_ICS}, and maximising with respect to the six values which parameterise the $P(M_r)$ curve in the case of the Naive model, and or $f_{\rm sfr}$ in the case of the Loudas25 model.
\begin{figure}
    \centering
    \includegraphics[width=\linewidth]{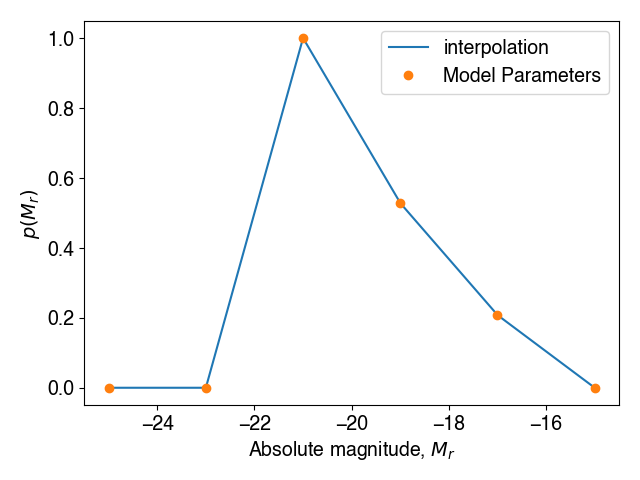}
    \caption{Fitted distribution of best-fit absolute magnitudes $M_r$ for the Naive model developed in this work, showing the model parameters, and the interpolation between them.}
    \label{fig:simple_abs_mags}
\end{figure}

\begin{figure}
    \centering
    \includegraphics[width=\linewidth]{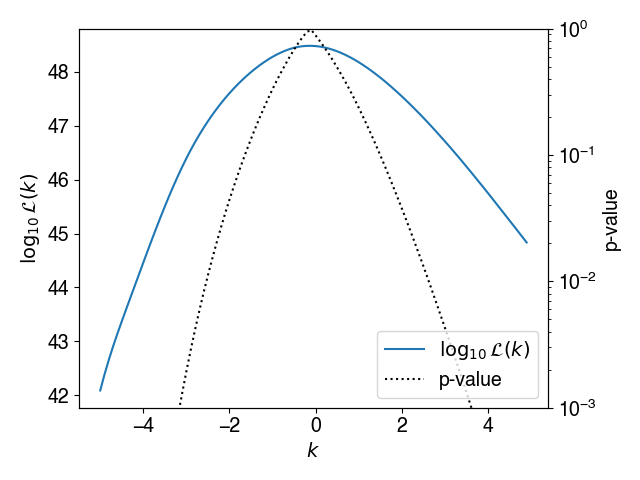}
    \caption{Fitted log likelihood of the Naive model (this work) as a function of $k$, $\log_{10} \mathcal{L}(k)$, while holding other model parameters fixed at their values given in Figure~\ref{fig:simple_abs_mags}.}
    \label{fig:kcorr}
\end{figure}

In the case of the Naive model, we find that our numerical procedure produces a peak in $P(M_r)$ near $M_r = -21$. Our preliminary investigation (see \ref{sec:binning}) found no probability of there being an excess population of very dim ($M_r \ge -15$) or very bright ($M_r < -25$) hosts. However, we find a significant tail towards high magnitude (low-luminosity) hosts. Our best-fit $K_r$-correction parameter is $k=-0.1$; fixing other model parameters, and varying $k$, produces the $\mathcal{L}(k)$ curve given in Figure~\ref{fig:kcorr}. Using Wilks' theorem, $-2 \log \mathcal{L}(k_{\rm best})/\mathcal{L}(k_{\rm true}) \sim \chi^2_1$, where $k_{\rm true}$ is the true value of $k$ and $k_{\rm best}$ is the best-fit value, for large sample sizes; this allows us to construct a 68\% confidence interval for $k$ of $(-1.4,1.3)$, i.e., we find no evidence for a $K_r$-correction to the FRB absolute host magnitude distribution. This is not surprising, given that our most likely hosts span only a small redshift range: $z \le0.55$ for all but one FRB, and most fall in the range $0.2 \le z \le 0.4$.

\begin{figure}
    \centering
    \includegraphics[width=\linewidth]{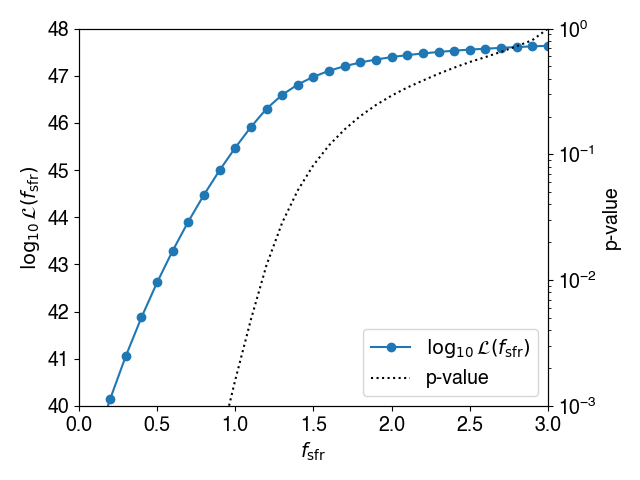}
    \caption{Fitted log likelihood of the Loudas25 model as a function of $f_{\rm sfr}$, $\log_{10} \mathcal{L}(f_{\rm sfr})$.}
    \label{fig:llsfr}
\end{figure}

In the case of the Loudas25 model, we show $\mathcal{L}(f_{\rm sfr})$ in Figure~\ref{fig:llsfr}. We obtain a best-fit value of $f_{\rm sfr}=3.0$, the highest value we investigate. Using Wilks' theorem as above, we find a 68\% lower bound on $f_{\rm sfr}$ to be 1.7; we exclude that $f_{\rm sfr} \le 1$ with a p-value of 0.12\%, i.e.\ at $3.0\,\sigma$ on a two-sided test. This observation is at odds with the scenario where a fraction of FRBs are stellar-mass tracers, as proposed by \citet{2026ApJ...996...78H}.

\begin{table}[]
    \centering
    \begin{tabular}{l|c c}
    Model & KS & $\log_{10} \mathcal{L}$ \\
    \hline
    Marnoch23                     & 0.078  & 48.35  \\
    Loudas25 ($f_{\rm sfr}=3.0$)  & 0.122  & 47.63  \\
    Naive (best fit)            & 0.079 &  48.48  \\
    \end{tabular}
    \caption{Value of the KS-like test statistic, and log-likelihood, for the three models considered in this work.}
    \label{tab:stats}
\end{table}

\begin{figure}
    \centering
    \includegraphics[width=\linewidth]{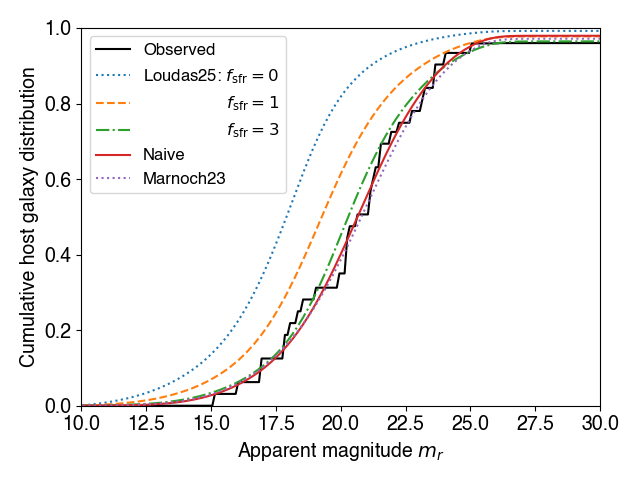}
    \caption{Cumulative observed (posterior) and prior distributions in $P(m_r)$ for ASKAP/CRAFT ICS FRBs according to the Loudas25 model, with $f_{\rm sfr} = 0$ (mass-weighted), $f_{\rm sfr} = 1$ (SFR-weighted), and $f_{\rm sfr} = 3.0$ (best fit); the Naive model developed in this work; and the Marnoch23 model. See text for explanation of calculation.}
    \label{fig:ks_like}
\end{figure}

We compare the goodness of fit between models both by comparing the peak fitted likelihoods, and by comparing the cumulative distributions (CDFs) of our priors and posteriors on $P(m_r)$, summed over all FRB host images. The cumulative distributions $C(m_r)$ predicted by the models are calculated as
\begin{eqnarray}
    C(m_r) & = & N_{\rm FRB}^{-1} \sum_{i=1}^{N_{\rm FRB}} \int_{-\infty}^{m_r} P(O|m_r^\prime) P_i(m_r^\prime) d m_r^\prime,
\end{eqnarray}
such that
\begin{eqnarray}
    C(m_r \to \infty) & = & N_{\rm FRB}^{-1} \sum_{i=1}^{N_{\rm FRB}} \left(1 - P_i(U)\right),
\end{eqnarray}
where $P_i$ indicates the probability for the $i^{\rm th}$ FRB. The cumulative observed distributions are generated from observed galaxy magnitudes $m_r$ weighted by posterior host probabilities $\POx$. Measuring the maximum differences between the CDFs produces a KS-like statistic \citep{kolmogorov,smirnov}. However, important exceptions prevent a quantitative interpretation: in these diagrams, both prior and posterior distributions vary with the model; each increment is not of equal value, but weighted by $P(O_i|\mathbf{x},N_O)$; and the normalisation of each scale is not unity, but is less, due to the probability of the true host being unseen. Nonetheless, smaller values of our KS-like statistic indicate a better fit. We show these values in Figure~\ref{fig:ks_like} for all three models, and in Table~\ref{tab:stats} give the values for the KS-like statistic, and maximum likelihood calculated above. The observed posterior (cumulative) distributions are almost identical, as discussed below in \S\,\ref{sec:updating}, and visually overlap each other.

Comparing both the KS-like test statistics and log-likelihood values, the Naive model gives a marginally better description of the data than the Marnoch23 model, with the Loudas25 model giving a less good fit. We should bear in mind that the Marnoch23, Loudas25, and Naive models have no, one, and seven degrees of freedom respectively, although the Marnoch23 model is explicitly fit to photometric data on many FRBs contained in the current sample, and so is not independent. We currently do not have a statistic by which we can compare likelihoods between models, so cannot assess whether or not the relative likelihoods, given the number of degrees of freedom, is evidence for or against any given model. Nonetheless, given that the Marnoch23 model has both fewer degrees of freedom and gives at least an equally good fit as other models, we suggest that this should be considered the default model.

\begin{figure}
    \centering
    \includegraphics[width=\linewidth]{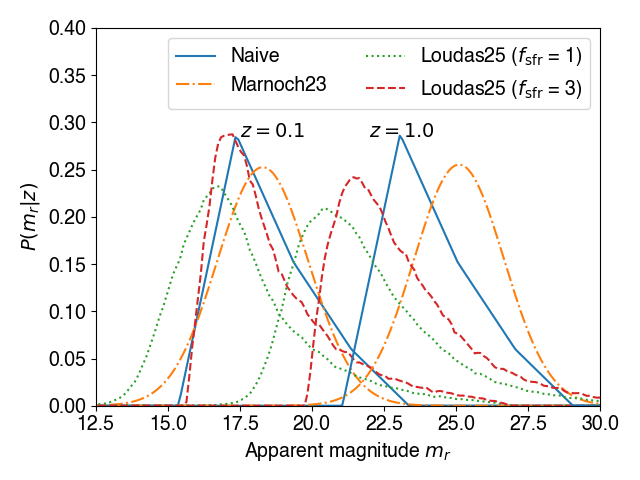}
    \caption{Apparent magnitude distributions, $P(m_r|z)$, produced by the empirical Marnoch23 model based on data from \citet{Marnoch2023}; the predictive Loudas25 models based on star-formation rate ($f_{\rm sfr}=1$), and the best-fit value ($f_{\rm sfr}=3$) from \citet{Loudas25}; and a Naive model developed in this work (see text).}
    \label{fig:pmrgz}
\end{figure}

We present the $P(m_r|z)$ distributions for the three models in Figure~\ref{fig:pmrgz}, using the best-fitting for the Naive and Loudas25 models, and also that for the Loudas25 model with $f_{\rm sfr} = 1$, at characteristic values of $z=0.1$ and $z=1.0$. Notably, the Loudas25 $f_{\rm sfr}=1$ model predicts FRB hosts to have much smaller magnitudes (i.e., be much brighter) than any of the models fit to data. By weighting the galaxy population selected by $\rm SFR ~(f_{sfr}=1)$, the Loudas25 model predicts brighter galaxies than unweighted galaxy populations. Clearly, the seemingly unphysical best-fit value of $f_{\rm sfr=3}$ is the Loudas25 model attempting to construct a $P(m_r)$ distribution with fainter galaxies than that predicted by SFR alone, while suppressing the abundance of brighter ones (which are likely associated with high-SFR, high-mass galaxies).

The difference between an SFR-weighted population and the best-fit models increases with redshift: at $z=0.1$, the three fitted distributions are very similar, whereas by $z=1.0$, the Loudas25, Naive, and Marnoch23 models produce increasingly faint distributions. Since only one FRB in our data sample \citep[FRB\,20220610A, at $z=1.016$ ][]{Ryder2023,2024ApJ...963L..34G} contains a probable host galaxy with measured magnitude at $z > 0.53$, and the model predictions diverge at high redshift, we urge caution in interpreting our results in the $z \gtrsim 0.5$ range.

\subsection{Updating FRB host galaxy predictions}
\label{sec:updating}

So far in this work, we have phrased our conclusions in terms of `what is the intrinsic magnitude distribution of FRB host galaxies?'. However, for those groups using PATH to identify the most likely FRB hosts, a much more relevant conclusion is: `how does this change our conclusions on true FRB hosts'?

\begin{figure}
    \centering
    \includegraphics[width=\linewidth]{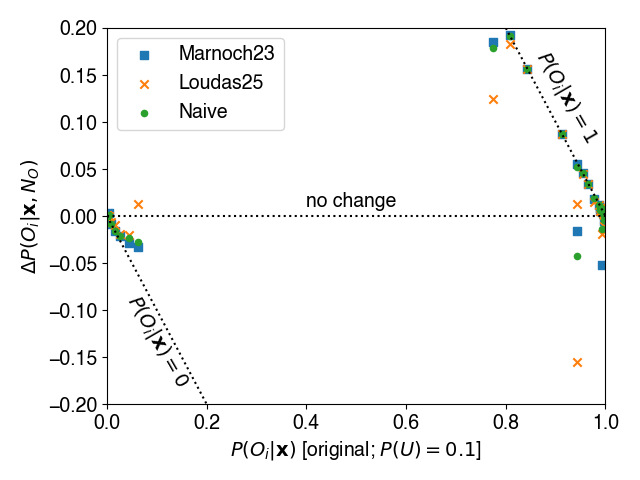}
    \caption{Change in posterior confidence $\Delta P(O_i|\mathbf{x},N_O)$ as a function of posterior values $P(O_i|\mathbf{x},N_O)$ calculated from the original PATH methodology, as per the Marnoch23, Loudas25, and Naive models with best-fit values. Positive values indicate increased confidence.}
    \label{fig:posteriors}
\end{figure}

In Figure~\ref{fig:posteriors}, we give the distribution of posterior $P(O_i|\mathbf{x})$ values for ICS FRBs using a standard PATH analysis. Like \citet{Shannon_ICS}, we use an offset prior with a scale factor of half of the galaxy's half-light radius. Unlike \citet{Shannon_ICS}, we use $P(U)=0.1$, due to the dangers of using $P(U)=0$ noted by \Andersen. We show the change in posterior values, $\Delta P(O_i|\mathbf{x},N_O)$, when using the Marnoch23 and best-fit Naive and Loudas25 $f_{\rm sfr}=3.0$ models. To guide the eye, we plot loci corresponding to the cases where the new posterior estimates $P(O_i|\mathbf{x})$ are zero and unity.
Reassuringly, all models give broadly similar results, identifying the same most probable hosts as those previously produced by PATH: we give updated values in Table~\ref{tab:path}. Furthermore, all three models produce near-identical posteriors which tend to increase our posterior likelihoods, with many previously likely host candidates being promoted to the $P(O_i|\mathbf{x})=1$ line, and many previously unlikely candidates being demoted to the $P(O_i|\mathbf{x})=0$ line. In particular, FRBs 20210807D and 20240210A,  which under a PATH analysis with $P(U)=0.1$ do not meet the usual criteria of $P(O|x) \ge 0.95$ for `firm' host galaxy associations (with most probable hosts having $P(O|x)=0.843$ and $0.718$ respectively), have $P(O|x)=1.00$ for all three models used here, validating the choice of $P(U)=0$ by \citet{Shannon_ICS}. FRB~20181112A previously had $P(O|x) =0.775$, with the Loudas25 model producing $P(O|x) = 0.896$, and the Marnoch23 and Naive models producing $P(O|x) = 0.960$ and $0.953$ respectively; given that the latter give better fits, we suggest this also be considered a firm host identification. Our updated model also confirms the result from \citet{Marnoch2023} that there are no viable candidates for FRB\,20210912A, with all models agreeing on $P(U|x)=1$. Finally, we confirm the host galaxy of FRB\,20190611B ($P(O|x) \ge 0.957$), the identification of which had been called into question by \citet{CordesTauRedshift2022} due to its relatively high redshift for its DM, and significant scattering value.

\subsection{Well-justified unseen priors}
\label{sec:unseen}

\begin{figure*}
    \centering
    \includegraphics[width=0.49\linewidth]{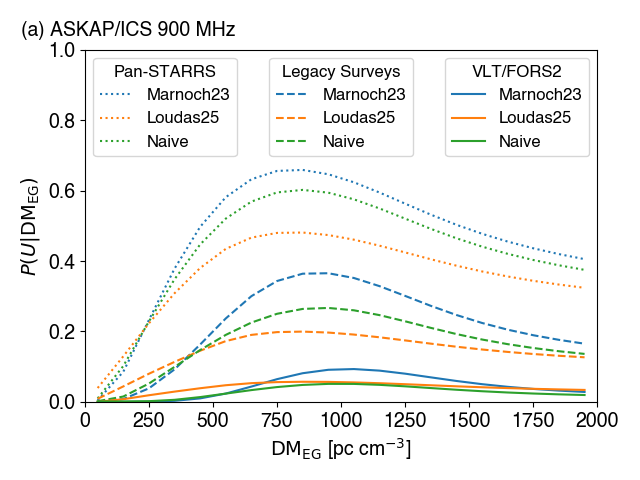} \includegraphics[width=0.49\linewidth]{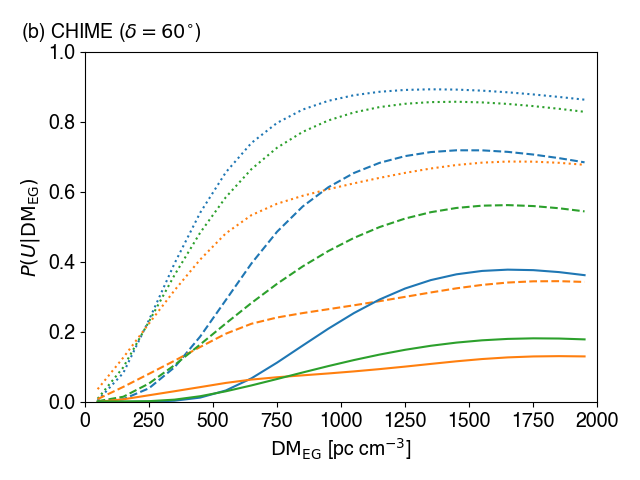}
    \includegraphics[width=0.49\linewidth]{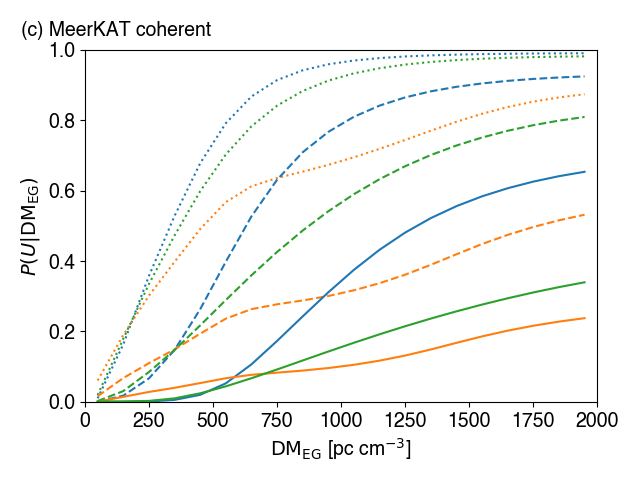} \includegraphics[width=0.49\linewidth]{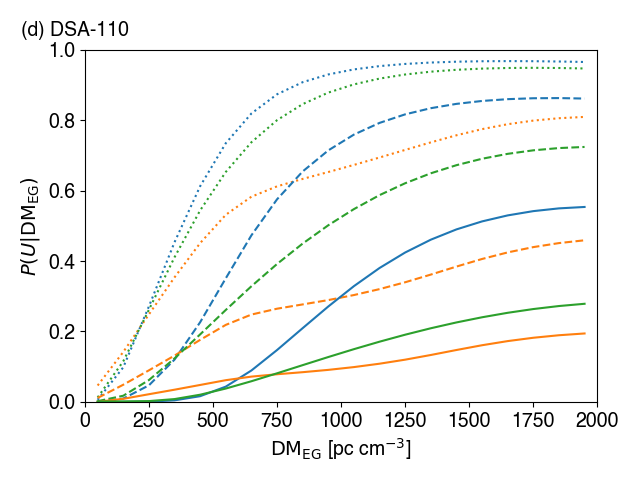}
    \caption{Estimated probabilities, $P(U|{\rm DM}_{\rm EG})$, of the true host being unseen in an optical image as a function of extragalactic DM, ${\rm DM}_{\rm EG}$, for each of the three host models developed in this work, and three optical follow-up sensitivities, shown for: (a) ASKAP ICS observations at 892 MHz; (b) CHIME observations at a declination of $\delta=60^{\circ}$; (c) MeerKAT coherent detection; and (d) DSA-110.}
    \label{fig:pu}
\end{figure*}

Previous uses of PATH have chosen a variety of values for the unseen prior, $P(U)$, ranging from 0 \citep[e.g.][]{Shannon_ICS} to $P(U)=0.2$ by \citet{DSA_Sharma_sfr}; only \citet{2025arXiv250801648C}, who also used the results of \citet{Marnoch2023}, have estimated $P(U)$ based on an assumed host galaxy distribution and FRB redshift.

Here, we can readily test these assumptions. In Figure~\ref{fig:pu} we plot $P(U|{\rm DM}_{\rm EG})$ as a function of ${\rm DM}_{\rm EG}$ for all three fit models (giving $P(m_r|{\rm DM_{\rm EG}})$) and optical depths (giving $P(O|m_r)$) used here, for the ASKAP/ICS observations at $\sim 900$\,MHz, CHIME observations at $\delta=60^{\circ}$, MeerKAT coherent detections, and DSA-110.

For some critical value of ${\rm DM}_{\rm EG}$, the function $P(z|{\rm DM}_{\rm EG})$ calculated by {\sc zDM} will predict a decreasing $z$ with increasing ${\rm DM}_{\rm EG}$. This is due to the probability of detecting a necessarily very bright FRB from a very large distance becoming less than the probability of a large over-fluctuation in DM from a nearby FRB. While the exact behaviour of this effect depends on the FRB luminosity function, frequency dependence, DM host contribution, and FRB population evolution \citep{james_zdm_2022}, it will occur at lower values of ${\rm DM}_{\rm EG}$ for less-sensitive instruments. The consequence is that FRBs with very high ${\rm DM}_{\rm EG}$ are predicted to have lower redshift and therefore brighter apparent magnitudes, i.e., they will be easier to detect despite their very high DM values.
Using the FRB population parameter fits of \citet{HoffmannHalo26}, this effect becomes significant in the ASKAP ICS survey for ${\rm DM}_{\rm EG} > 1000$\,\pccc, and $>1500$\,\pccc\ for CHIME, but is not relevant for ${\rm DM}_{\rm EG} < 2000$\,\pccc\ for MeerKAT or DSA-100. Different models for the $P(m_r|z)$ distribution also produce significantly different predictions for ${\rm DM}_{\rm EG} \gtrsim 500$. Combined with the aforementioned uncertainties in the $P(z|{\rm DM}_{\rm EG})$,  we again highlight that the predictions from this work should be treated with caution in this regime.

With these caveats, we observe that $P(U)$ is highly DM-dependent, such that setting $P(U)$ constant for an FRB survey with a large range of FRB DMs (as is usually performed in the literature) is ill-advised. In the case of surveys such as Pan-STARRS, $P(U)$ will reach 50\% by a relatively modest ${\rm DM}_{\rm EG} = 500$\,\pccc, while observations with VLT/FORS2 produce $P(U)\lesssim 0.05$ in this range. Thus we recommend re-running PATH on FRB samples from existing surveys, e.g., the DSA FRB samples of \citet{DSA_Sharma_sfr} and \citet{2025NatAs...9.1226C}, or the MeerKAT FRBs of \citet{2023MNRAS.524.4275J} and \citet{2026MNRAS.545f2144P}, to generate more-robust host associations. Due to its simplicity, we recommend using the Marnoch23 model to generate PATH priors. However, a more thorough approach would use all three models, and treat the variation in host galaxy posteriors between them as a model systematic.
This methodology should also be relevant to localisations with CHIME outriggers \citep{2025ApJS..280....6C}, and perhaps very low-DM CHIME FRBs detected without baseband data \citep[e.g., ][]{2024ApJ...971L..51B}.

\subsection{Accounting for redshift}
\label{sec:redshift}

\begin{figure}
    \centering
    \includegraphics[width=\linewidth]{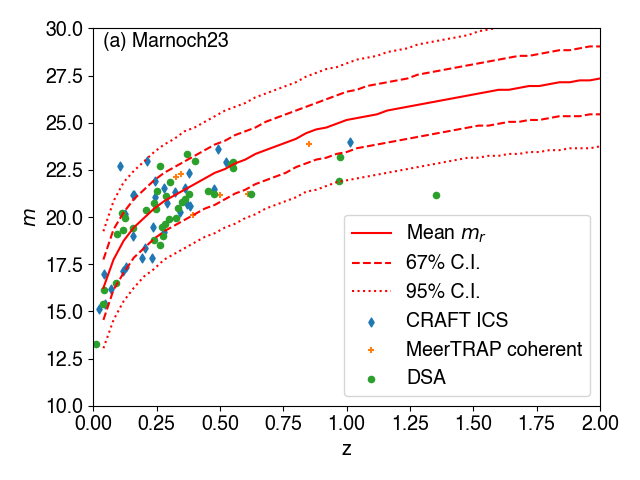}
    \includegraphics[width=\linewidth]{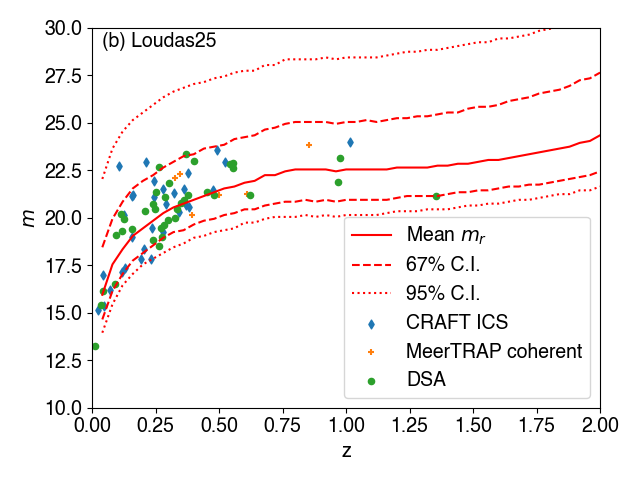}
    \includegraphics[width=\linewidth]{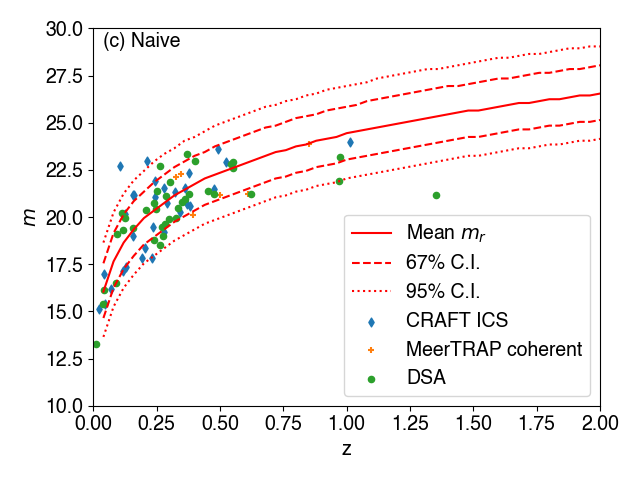}
    \caption{Mean, and 67\% ($\pm 1 \sigma$) and 95\% confidence ($\pm 2 \sigma$) intervals, of the FRB $r$-band host galaxy magnitude distribution, $m_r$, as a function of redshift, using Marnoch23, Loudas25, and Naive models. Also shown are observed magnitudes for probable FRB host galaxies (measured using a variety of filters) for FRBs detected in CRAFT ICS \citep{Shannon_ICS}, MeerTRAP coherent \citep{2026MNRAS.545f2144P}, and DSA \citep{DSA_Sharma_sfr,2025NatAs...9.1226C} observations.}
    \label{fig:zmr}
\end{figure}

Our updates to PATH assume that an optical image is being analysed with no redshift information available. This is because PATH was designed to identify a most likely host for further spectroscopic follow-up, meaning that no spectroscopic information should be assumed when running it. Furthermore, when performing FRB follow-up, spectroscopic data is usually obtained only for the most probable host galaxy for that FRB. 

In a future work, we aim to develop methods to incorporate partial redshift information into our likelihood estimates. For now, we show the predicted $P(m_r|z)$ distributions of FRB hosts for each of our models in Figure~\ref{fig:zmr}, and compare these to measured redshifts of FRB hosts from major surveys. While it appears that the high-redshift FRB hosts identified by MeerTRAP and DSA are brighter than those predicted by the Marnoch23 and Naive models, and agree better with the Loudas25 model, we note that both samples are incomplete, with only those FRBs with detectable hosts being published (the majority of these measurements were taken with an $r$-band filter, so we discount this as a source of discrepancies). A counter-example is FRB\,20240304B with ${\rm DM}_{\rm EG}=2390$\,\pccc, where \citet{2025arXiv250801648C} report a relatively faint most probable host at $z=2.148$, having $m_{\rm F200W} = 27.82$ and $m_{\rm 322W2} = 28.72$. This emphasises another source of bias not yet discussed in this work (reporting bias), and hence the importance of publishing all FRB detections, and optical follow-up data, irrespective of whether or not a probable host has been identified.

\section{Discussion and conclusion}
\label{sec:discussion}

We have extended the PATH framework to allow for models of expected galaxy properties to be incorporated into FRB host identification in optical images. This framework allows for a robust, and physically consistent, method of determining priors that an FRB host galaxy is unseen in optical images with arbitrary magnitude limits, thereby allowing data from an increased range of experiments to be used in both FRB host galaxy and population studies, and systematics due to unseen hosts to be explicitly accounted for. 

We have considered three models for the true distribution of FRB host galaxy $r$-band magnitudes: the `Marnoch23' model of \citet{2026arXiv260300371J} based on \citet{Marnoch2023}, which is an empirical fit to known FRB hosts; the `Loudas25' model based on \citet{Loudas25}, which weights the galaxy population by a linear combination of stellar mass and star-formation rate; and the `Naive' model based on absolute galaxy magnitudes, developed in this work. The parameters of the latter two are fit to the observed host galaxy properties of 32 FRBs detected by ASKAP/ICS observations.

Incorporating these models into PATH, we recalculated the probabilities of candidate galaxies being the true host for FRBs detected by the ASKAP/CRAFT ICS system. We find that all three models tend to increase the confidence in previously-identified most likely hosts. Using the Loudas25 model, we find strong evidence (p-value 0.86\%) that the true distribution of FRB host galaxies is fainter than that predicted by star-formation, and overwhelming evidence (p-value $10^{-9}$) that it is fainter than that of a stellar-mass weighting. Using the Marnoch23 and Naive models, both of which give more likely fits, predict an even fainter distribution of FRB hosts, especially at higher redshifts. Due to its simplicity, we recommend using the Marnoch23 model to generate FRB host galaxy priors within the PATH framework, although using the Loudas25 and Naive models will help to evaluate model systematics.

We consider two broad explanations regarding the relative faintness of the FRB host distribution. Firstly, this effect may be due to e.g.\ a metallicity preference in FRB hosts, as suggested by \citet{DSA_Sharma_sfr}. However, here we invoke it due to an {\it excess} of dim FRB hosts, rather than a deficit. It is in principle possible to extend the Loudas25 model to a 2D distribution in ($\rm SFR$, $\rm M^*$) space, which would then include metallicity either implicitly or explicitly. We might also weight by star-formation history, e.g.\ via an exponential time-delay representing merger scenarios, by building in a model of each galaxy's evolution.

Secondly, another possible explanation is that the apparent excess of dim hosts is in fact a deficit of bright ones caused by FRB selection effects. FRB detection experiments are known to be biased against highly scattered FRBs, since scattering smears the burst fluence (i.e.\ the signal) over a longer time-duration (i.e.\ more noise) \citep{Cordes_McLaughlin_2003}. Larger host galaxies would be expected to result in more-scattered FRBs, potentially due to having more turbulent, ionised gas, but also due to a larger geometric lever-arm between the FRB location and the host interstellar medium. This would then reduce the rate of FRBs detected from these galaxies, similarly to the rate reduction viewed by CHIME through the plane of the Milky Way \citep{CHIMECat2} and the Cygnus region \citep{2026ApJ...997L...5P}. A correlation between galaxy inclination and scattering has been reported by \citet{2024Natur.634.1065B}, which is consistent with such a scattering-bias effect, although there is no direct evidence for a correlation between FRB scattering properties and other FRB host properties such as stellar mass \citep{2025PASA...42..157G}.

Given that most of the data used to construct our models were from the relatively nearby Universe ($z \lesssim 0.5$), our host models begin to diverge at high redshifts, and we do not consider our predictions reliable for $z \ge 1$. Of great benefit would be incorporating fits to data from DSA-110 and MeerTRAP, which probe to higher redshifts than ASKAP, and would allow sufficient data to properly fit for the extended parameter space. We have also made several approximations when scaling observed galaxy magnitudes in other bands to r-band magnitudes, in which our models are specified. A more complete model would directly estimate magnitudes in multiple filters. Another extension of this work would be to enable the {\sc zDM} code, which we use to predict the redshift distributions of FRB hosts, to accept host galaxy candidate posterior probabilities, and calculate updated redshift distributions for FRBs with unseen host galaxies. This would allow the incorporation of data where the true FRB host is too faint to detect in optical observations, or where the identification of the true host is uncertain. Doing so would both allow and require the FRB host galaxy distribution, and FRB population, to be simultaneously fit, which is a complex task that we leave to a future work.

For now, we encourage the updated version of PATH to be used for FRB host galaxy identification, and have made available code allowing the use of each model.

\begin{acknowledgement}
We would like to thank the anonymous referee for their constructive feedback that helped us improve the clarity of this work.
NL acknowledges support from the Venture Forward Fellowship of the Department of Astrophysical Sciences at Princeton University.

\end{acknowledgement}

\paragraph{Funding Statement}

The authors declare no external sources of funding.

\paragraph{Competing Interests}

The authors declare no competing interests.

\paragraph{Data Availability Statement}

The data and results from this work can be obtained from the \zdm\ GitHub codebase \citep{zdm}, at \url{https://github.com/FRBs/zdm}.

\printendnotes

\printbibliography

\appendix

\input{path_table}

\section{Binning in the Naive model}
\label{sec:binning}

Our Naive model, described in \S~\ref{sec:naive}, is parameterised by an arbitrary number of bins, $N_{\rm bins}$, in absolute magnitude space. There is no a~priori prescription for the `correct' number of bins, or the correct minimum and maximum absolute magnitudes to use. Good statistical practice suggests that the total number of bins remains significantly smaller than the number of FRBs we analyse here (32), while we wish to retain sufficiently many to allow the model to be flexible enough to describe the (unknown) FRB host galaxy distribution. Here we test the effects of this arbitrariness, to determine how robust our conclusions are against variations in these parameters.

\begin{figure}
    \centering
    \includegraphics[width=\linewidth]{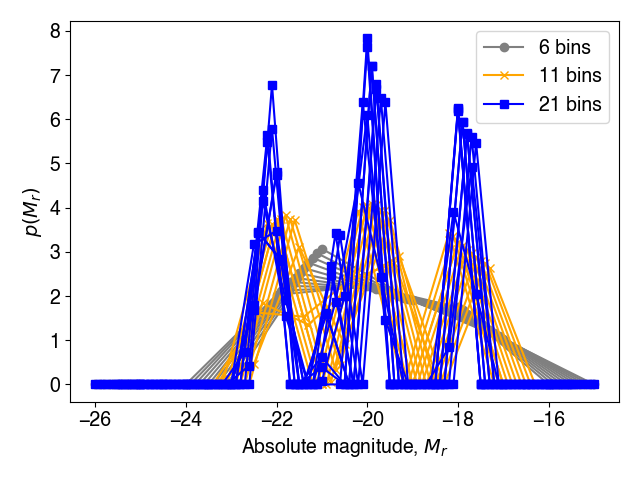}
    \caption{Best fitting distributions $P(M_r)$ found for various combinations of $N_{\rm bins}$, $M_r^{\rm min}$, and $M_r^{\rm max}$.}
    \label{fig:mrsys}
\end{figure}

We first tried $N_{\rm bins} = 10$, the range $M_r^{\rm max}=-30$, $M_r^{\rm min}=-10$, and the fitting procedure of \S~\ref{sec:fitting}, and found that the model never predicts any probability of $M_r<-25$ or $M_r > -15$. We then investigated the effects of bin location and number by varying $M_r^{\rm min}$ from $-26$ to $-25$ in increments of $0.1$, setting $M_r^{\rm max} = M_r^{\rm min}+10$, and consider $N_{\rm bins}=\{ 6,11,21 \}$ (i.e., bin spacings of 2, 1, and 0.5 magnitudes). Running these 33 scenarios produced the set of best-fit $M_r$ distributions shown in Figure~\ref{fig:mrsys}. We find that as $N_{\rm bins}$ increases, the model predicts more spiky structures in $P(M_r)$, which we interpret as the model over-fitting the data. We find the best-fit log-likelihood to increase with $N_{\rm bins}$, as expected for greater degrees of freedom, and to vary with exact placement by $\pm 0.1$.

We can use Wilks' theorem to determine the significance of increasing the number of bins, since halving the bin spacing preserves the locations of the original bins in magnitude space, i.e., it produces a nested model. Doing so shows that the increased likelihood with increasing bins is negligible, and would occur at least 96\% of the time under the null hypothesis that the 6-bin model correctly describes the data. While Wilks' theorem is only valid as the number of data points tends to infinity, our complete lack of evidence for a significant increase in explanatory power with increased model bins makes us confident in our conclusion. We therefore use six bins for our analysis. We consider that fewer than six bins results in insufficient flexibility, which is the purpose of this model.

\end{document}

%% file: path_table.tex
\begin{table*}
\caption{Updated FRB PATH associations\label{tab:path}, giving priors and posteriors for the traditional PATH analysis with $P(U)=0.1$, and our three models for $p(m)$, using best-fit values for the Loudas and naive models (see text). Only galaxies with \POx $> 10^{-4}$ in at least one model are reported. Note that \PO\ values in the original PATH analysis include the $\Sigma_j$ values, but have been renormalised according to Eq.~\ref{eq:orig_path}, while other $P(O)$ values do not include the equivalent factor, $\rho(m)$.}
\begin{tabular}{ccccc|cc|cc|cc|cc|}
\multicolumn{5}{c|}{TNS Name}       & \multicolumn{2}{c|}{Original PATH} & \multicolumn{2}{c|}{Marnoch23} & \multicolumn{2}{c|}{Loudas25} & \multicolumn{2}{c}{Naive} \\
RA$_{\rm cand}$ & Dec$_{\rm cand}$ & $\theta$ & \halflight  & mag & $P(U)$ & $P(U|\mathbf{x})$ & $P(U)$ & $P(U|\mathbf{x})$ & $P(U)$ & $P(U|\mathbf{x})$ & $P(U)$ & $P(U|\mathbf{x})$ \\
 (deg) & (deg) & ($''$) & ($''$) &     & \PO & \POx  & \PO & \POx  & \PO & \POx  & \PO & \POx \\
\hline
\hline
\multicolumn{5}{c|}{FRB20180924B} & 0.100 &  0.002 & 0.001 &  0.000 & 0.024 &  0.000 & 0.003 &  0.000 \\ 
\hline
326.1052 & -40.9002 & 0.79 & 1.00 & 19.98 &0.870 &  0.998 & 0.197 &  1.000 & 0.163 &  1.000 & 0.191 &  1.000 \\ 
\hline
\multicolumn{5}{c|}{FRB20181112A} & 0.100 &  0.138 & 0.026 &  0.007 & 0.054 &  0.019 & 0.029 &  0.008 \\ 
\hline
327.3486 & -52.9709 & 0.40 & 0.68 & 21.58 &0.078 &  0.775 & 0.195 &  0.960 & 0.145 &  0.900 & 0.200 &  0.953 \\ 
327.3496 & -52.9696 & 5.36 & 1.06 & 19.22 &0.757 &  0.061 & 0.085 &  0.029 & 0.175 &  0.074 & 0.103 &  0.034 \\ 
327.3467 & -52.9704 & 4.24 & 0.28 & 25.67 &0.003 &  0.026 & 0.008 &  0.004 & 0.009 &  0.007 & 0.009 &  0.005 \\ 
327.3483 & -52.9729 & 7.03 & 0.62 & 22.06 &0.051 &  0.000 & 0.198 &  0.000 & 0.119 &  0.000 & 0.172 &  0.000 \\ 
\hline
\multicolumn{5}{c|}{FRB20190102C} & 0.100 &  0.003 & 0.001 &  0.000 & 0.022 &  0.002 & 0.003 &  0.000 \\ 
\hline
322.4153 & -79.4758 & 0.34 & 0.87 & 21.25 &0.744 &  0.997 & 0.145 &  1.000 & 0.087 &  0.998 & 0.115 &  1.000 \\ 
\hline
\multicolumn{5}{c|}{FRB20190608B} & 0.100 &  0.034 & 0.001 &  0.000 & 0.022 &  0.000 & 0.003 &  0.000 \\ 
\hline
334.0204 & -7.8988 & 2.87 & 1.66 & 17.15 &0.893 &  0.966 & 0.092 &  1.000 & 0.107 &  1.000 & 0.096 &  1.000 \\ 
\hline
\multicolumn{5}{c|}{FRB20190611B} & 0.100 &  0.031 & 0.000 &  0.001 & 0.016 &  0.035 & 0.002 &  0.003 \\ 
\hline
320.7428 & -79.3973 & 2.15 & 0.56 & 23.09 &0.334 &  0.943 & 0.041 &  0.999 & 0.041 &  0.956 & 0.049 &  0.996 \\ 
320.7440 & -79.3984 & 3.28 & 0.31 & 25.60 &0.048 &  0.016 & 0.001 &  0.000 & 0.012 &  0.007 & 0.002 &  0.001 \\ 
320.7497 & -79.3972 & 3.13 & 0.28 & 26.76 &0.022 &  0.008 & 0.000 &  0.000 & 0.001 &  0.002 & 0.000 &  0.000 \\ 
320.7383 & -79.3976 & 4.85 & 0.39 & 24.28 &0.127 &  0.000 & 0.009 &  0.000 & 0.025 &  0.000 & 0.019 &  0.000 \\ 
\hline
\multicolumn{5}{c|}{FRB20190711A} & 0.100 &  0.014 & 0.023 &  0.059 & 0.053 &  0.189 & 0.027 &  0.080 \\ 
\hline
329.4195 & -80.3581 & 1.43 & 0.58 & 24.37 &0.409 &  0.942 & 0.073 &  0.927 & 0.043 &  0.788 & 0.060 &  0.900 \\ 
329.4121 & -80.3571 & 4.58 & 0.56 & 25.45 &0.190 &  0.043 & 0.013 &  0.015 & 0.015 &  0.023 & 0.015 &  0.020 \\ 
\hline
\multicolumn{5}{c|}{FRB20190714A} & 0.100 &  0.003 & 0.011 &  0.000 & 0.043 &  0.000 & 0.016 &  0.000 \\ 
\hline
183.9795 & -13.0211 & 0.81 & 0.93 & 19.56 &0.707 &  0.997 & 0.135 &  1.000 & 0.225 &  1.000 & 0.174 &  1.000 \\ 
\hline
\multicolumn{5}{c|}{FRB20191001A} & 0.100 &  0.006 & 0.009 &  0.000 & 0.039 &  0.000 & 0.013 &  0.000 \\ 
\hline
323.3519 & -54.7485 & 0.97 & 1.35 & 17.81 &0.457 &  0.993 & 0.065 &  0.999 & 0.079 &  0.999 & 0.066 &  0.999 \\ 
323.3486 & -54.7482 & 6.49 & 1.45 & 17.83 &0.443 &  0.001 & 0.066 &  0.001 & 0.080 &  0.001 & 0.067 &  0.001 \\ 
\hline
\multicolumn{5}{c|}{FRB20191228A} & 0.100 &  0.192 & 0.000 &  0.000 & 0.016 &  0.009 & 0.002 &  0.001 \\ 
\hline
344.4308 & -29.5943 & 1.12 & 0.31 & 22.53 &0.011 &  0.808 & 0.037 &  1.000 & 0.040 &  0.991 & 0.046 &  0.999 \\ 
\hline
\multicolumn{5}{c|}{FRB20200430A} & 0.100 &  0.003 & 0.002 &  0.000 & 0.027 &  0.001 & 0.005 &  0.000 \\ 
\hline
229.7064 & 12.3766 & 1.07 & 0.69 & 20.69 &0.826 &  0.997 & 0.186 &  1.000 & 0.126 &  0.999 & 0.161 &  1.000 \\ 
\hline
\multicolumn{5}{c|}{FRB20200906A} & 0.100 &  0.004 & 0.019 &  0.000 & 0.047 &  0.001 & 0.022 &  0.000 \\ 
\hline
53.4958 & -14.0832 & 1.12 & 1.35 & 19.63 &0.879 &  0.996 & 0.122 &  1.000 & 0.211 &  0.999 & 0.154 &  1.000 \\ 
\hline
\multicolumn{5}{c|}{FRB20210117A} & 0.100 &  0.004 & 0.076 &  0.009 & 0.071 &  0.014 & 0.055 &  0.008 \\ 
\hline
339.9795 & -16.1515 & 0.91 & 0.51 & 22.95 &0.512 &  0.996 & 0.193 &  0.991 & 0.114 &  0.986 & 0.158 &  0.992 \\ 
\hline
\multicolumn{5}{c|}{FRB20210320C} & 0.100 &  0.002 & 0.002 &  0.000 & 0.025 &  0.000 & 0.004 &  0.000 \\ 
\hline
204.4589 & -16.1226 & 0.47 & 1.02 & 19.23 &0.813 &  0.998 & 0.175 &  1.000 & 0.212 &  1.000 & 0.198 &  1.000 \\ 
\hline
\multicolumn{5}{c|}{FRB20210807D} & 0.100 &  0.157 & 0.000 &  0.000 & 0.001 &  0.000 & 0.001 &  0.000 \\ 
\hline
299.2201 & -0.7623 & 4.83 & 2.32 & 17.35 &0.671 &  0.843 & 0.160 &  1.000 & 0.120 &  1.000 & 0.140 &  1.000 \\ 
\hline
\multicolumn{5}{c|}{FRB20210912A} & 0.100 &  1.000 & 0.273 &  1.000 & 0.096 &  1.000 & 0.134 &  1.000 \\ 
\hline
\hline
\multicolumn{5}{c|}{FRB20211127I} & 0.100 &  0.088 & 0.000 &  0.000 & 0.007 &  0.000 & 0.001 &  0.000 \\ 
\hline
199.8082 & -18.8379 & 2.18 & 5.07 & 15.38 &0.245 &  0.912 & 0.100 &  1.000 & 0.101 &  1.000 & 0.100 &  1.000 \\ 
199.8080 & -18.8402 & 8.78 & 2.28 & 18.34 &0.008 &  0.000 & 0.185 &  0.000 & 0.159 &  0.000 & 0.175 &  0.000 \\ 
\hline
\multicolumn{5}{c|}{FRB20211203C} & 0.100 &  0.002 & 0.024 &  0.000 & 0.050 &  0.000 & 0.026 &  0.000 \\ 
\hline
204.5626 & -31.3801 & 0.54 & 0.89 & 20.25 &0.789 &  0.998 & 0.112 &  1.000 & 0.204 &  1.000 & 0.141 &  1.000 \\ 
\hline
\multicolumn{5}{c|}{FRB20211212A} & 0.100 &  0.009 & 0.000 &  0.000 & 0.005 &  0.000 & 0.001 &  0.000 \\ 
\hline
157.3509 & 1.3608 & 1.45 & 2.72 & 16.21 &0.900 &  0.991 & 0.154 &  1.000 & 0.171 &  1.000 & 0.167 &  1.000 \\ 
\hline
\end{tabular} 
\end{table*}

\begin{table*}
\caption{FRB PATH associations (continued).}
\begin{tabular}{ccccc|cc|cc|cc|cc|}
\multicolumn{5}{c|}{TNS Name}       & \multicolumn{2}{c|}{Original PATH} & \multicolumn{2}{c|}{Marnoch23} & \multicolumn{2}{c|}{Loudas25} & \multicolumn{2}{c}{Naive} \\
RA$_{\rm cand}$ & Dec$_{\rm cand}$ & $\theta$ & \halflight  & mag & $P(U)$ & $P(U|\mathbf{x})$ & $P(U)$ & $P(U|\mathbf{x})$ & $P(U)$ & $P(U|\mathbf{x})$ & $P(U)$ & $P(U|\mathbf{x})$ \\
 (deg) & (deg) & ($''$) & ($''$) &     & \PO & \POx  & \PO & \POx  & \PO & \POx  & \PO & \POx \\
\hline
\hline
\multicolumn{5}{c|}{FRB20220105A} & 0.100 &  0.023 & 0.027 &  0.004 & 0.054 &  0.008 & 0.029 &  0.004 \\ 
\hline
208.8038 & 22.4665 & 1.83 & 0.90 & 21.53 &0.512 &  0.977 & 0.168 &  0.996 & 0.188 &  0.992 & 0.213 &  0.996 \\ 
\hline
\multicolumn{5}{c|}{FRB20220501C} & 0.100 &  0.001 & 0.005 &  0.000 & 0.034 &  0.000 & 0.009 &  0.000 \\ 
\hline
352.3792 & -32.4907 & 0.30 & 0.90 & 20.57 &0.900 &  0.999 & 0.192 &  1.000 & 0.158 &  1.000 & 0.194 &  1.000 \\ 
\hline
\multicolumn{5}{c|}{FRB20220610A} & 0.100 &  0.007 & 0.294 &  0.059 & 0.102 &  0.026 & 0.145 &  0.021 \\ 
\hline
351.0735 & -33.5137 & 0.76 & 1.02 & 23.96 &0.250 &  0.993 & 0.133 &  0.941 & 0.109 &  0.974 & 0.191 &  0.979 \\ 
\hline
\multicolumn{5}{c|}{FRB20220725A} & 0.100 &  0.003 & 0.000 &  0.000 & 0.015 &  0.000 & 0.002 &  0.000 \\ 
\hline
353.3154 & -35.9903 & 0.50 & 1.77 & 17.83 &0.898 &  0.997 & 0.126 &  1.000 & 0.154 &  1.000 & 0.139 &  1.000 \\ 
\hline
\multicolumn{5}{c|}{FRB20220918A} & 0.100 &  0.011 & 0.045 &  0.002 & 0.063 &  0.005 & 0.040 &  0.003 \\ 
\hline
17.5917 & -70.8114 & 0.49 & 0.45 & 23.60 &0.069 &  0.989 & 0.188 &  0.998 & 0.104 &  0.995 & 0.140 &  0.997 \\ 
\hline
\multicolumn{5}{c|}{FRB20221106A} & 0.100 &  0.008 & 0.001 &  0.000 & 0.022 &  0.000 & 0.003 &  0.000 \\ 
\hline
56.7045 & -25.5696 & 1.33 & 2.50 & 18.34 &0.850 &  0.988 & 0.117 &  0.993 & 0.154 &  0.996 & 0.130 &  0.995 \\ 
56.7057 & -25.5701 & 3.24 & 1.07 & 21.07 &0.050 &  0.004 & 0.184 &  0.007 & 0.124 &  0.003 & 0.155 &  0.005 \\ 
\hline
\multicolumn{5}{c|}{FRB20230526A} & 0.100 &  0.002 & 0.000 &  0.000 & 0.017 &  0.000 & 0.002 &  0.000 \\ 
\hline
22.2326 & -52.7175 & 0.51 & 0.78 & 21.15 &0.490 &  0.998 & 0.157 &  1.000 & 0.099 &  1.000 & 0.126 &  1.000 \\ 
\hline
\multicolumn{5}{c|}{FRB20230708A} & 0.100 &  0.008 & 0.002 &  0.000 & 0.027 &  0.001 & 0.005 &  0.000 \\ 
\hline
303.1155 & -55.3563 & 0.14 & 0.58 & 22.73 &0.079 &  0.992 & 0.119 &  1.000 & 0.072 &  0.999 & 0.096 &  1.000 \\ 
\hline
\multicolumn{5}{c|}{FRB20230902A} & 0.100 &  0.002 & 0.004 &  0.000 & 0.033 &  0.001 & 0.008 &  0.000 \\ 
\hline
52.1400 & -47.3335 & 0.54 & 0.68 & 21.52 &0.586 &  0.998 & 0.192 &  1.000 & 0.136 &  0.999 & 0.175 &  1.000 \\ 
\hline
\multicolumn{5}{c|}{FRB20231226A} & 0.100 &  0.012 & 0.001 &  0.000 & 0.019 &  0.000 & 0.002 &  0.000 \\ 
\hline
155.3639 & 6.1097 & 1.99 & 1.79 & 19.01 &0.845 &  0.988 & 0.161 &  1.000 & 0.209 &  1.000 & 0.185 &  1.000 \\ 
\hline
\multicolumn{5}{c|}{FRB20240201A} & 0.100 &  0.046 & 0.001 &  0.000 & 0.025 &  0.000 & 0.004 &  0.000 \\ 
\hline
149.9062 & 14.0888 & 3.76 & 5.00 & 16.92 &0.900 &  0.954 & 0.052 &  1.000 & 0.049 &  1.000 & 0.049 &  1.000 \\ 
\hline
\multicolumn{5}{c|}{FRB20240210A} & 0.100 &  0.282 & 0.000 &  0.000 & 0.014 &  0.000 & 0.002 &  0.000 \\ 
\hline
8.7770 & -28.2721 & 9.42 & 6.42 & 15.13 &0.758 &  0.718 & 0.037 &  1.000 & 0.033 &  1.000 & 0.034 &  1.000 \\ 
\hline
\multicolumn{5}{c|}{FRB20240304A} & 0.100 &  0.008 & 0.034 &  0.003 & 0.055 &  0.003 & 0.032 &  0.002 \\ 
\hline
136.3305 & -16.1662 & 1.84 & 0.97 & 21.08 &0.900 &  0.992 & 0.137 &  0.997 & 0.205 &  0.997 & 0.183 &  0.998 \\ 
\hline
\multicolumn{5}{c|}{FRB20240310A} & 0.100 &  0.002 & 0.023 &  0.000 & 0.050 &  0.000 & 0.025 &  0.000 \\ 
\hline
17.6219 & -44.4393 & 0.39 & 1.06 & 20.16 &0.687 &  0.997 & 0.109 &  0.999 & 0.201 &  0.999 & 0.137 &  0.999 \\ 
17.6228 & -44.4387 & 3.66 & 0.93 & 21.80 &0.141 &  0.001 & 0.177 &  0.001 & 0.160 &  0.001 & 0.201 &  0.001 \\ 
\end{tabular} 
\end{table*}

%% file: bibliography.bib
@ARTICLE{CordesTauRedshift2022,
       author = {{Cordes}, J.~M. and {Ocker}, Stella Koch and {Chatterjee}, Shami},
        title = "{Redshift Estimation and Constraints on Intergalactic and Interstellar Media from Dispersion and Scattering of Fast Radio Bursts}",
      journal = {\apj},
         year = 2022,
        month = jun,
       volume = {931},
       number = {2},
          eid = {88},
        pages = {88},
          doi = {10.3847/1538-4357/ac6873},
archivePrefix = {arXiv},
       eprint = {2108.01172},
 primaryClass = {astro-ph.HE},
       adsurl = {https://ui.adsabs.harvard.edu/abs/2022ApJ...931...88C}
}

@misc{zdm,
  author = {{James}, C. W. and {Prochaska}, J. X. and {Ghosh}, E. M.},
  title = {zdm},
  publisher = {GitHub},
  journal = {GitHub repository},
  howpublished = {\url{https://zenodo.org/record/5213780\#.YRxh5BMzZKA}},
  version = {0.1},
  date = {2021-08-18},
  year = 2021
  }

@ARTICLE{2020ApJ...903..152H,
       author = {{Heintz}, Kasper E. and {Prochaska}, J. Xavier and {Simha}, Sunil and {Platts}, Emma and {Fong}, Wen-fai and {Tejos}, Nicolas and {Ryder}, Stuart D. and {Aggerwal}, Kshitij and {Bhandari}, Shivani and {Day}, Cherie K. and {Deller}, Adam T. and {Kilpatrick}, Charles D. and {Law}, Casey J. and {Macquart}, Jean-Pierre and {Mannings}, Alexandra and {Marnoch}, Lachlan J. and {Sadler}, Elaine M. and {Shannon}, Ryan M.},
        title = "{Host Galaxy Properties and Offset Distributions of Fast Radio Bursts: Implications for Their Progenitors}",
      journal = {\apj},
         year = 2020,
        month = nov,
       volume = {903},
       number = {2},
          eid = {152},
        pages = {152},
          doi = {10.3847/1538-4357/abb6fb},
archivePrefix = {arXiv},
       eprint = {2009.10747},
 primaryClass = {astro-ph.GA},
       adsurl = {https://ui.adsabs.harvard.edu/abs/2020ApJ...903..152H}
}

@article{planck_collaboration_planck_2016,
       author = {{Planck Collaboration} and {Ade}, P.~A.~R. and {Aghanim}, N. and {Arnaud}, M. and {Ashdown}, M. and {Aumont}, J. and {Baccigalupi}, C. and {Banday}, A.~J. and {Barreiro}, R.~B. and {Bartlett}, J.~G. and {Bartolo}, N. and {Battaner}, E. and {Battye}, R. and {Benabed}, K. and {Beno{\^\i}t}, A. and {Benoit-L{\'e}vy}, A. and {Bernard}, J. -P. and {Bersanelli}, M. and {Bielewicz}, P. and {Bock}, J.~J. and {Bonaldi}, A. and {Bonavera}, L. and {Bond}, J.~R. and {Borrill}, J. and {Bouchet}, F.~R. and {Boulanger}, F. and {Bucher}, M. and {Burigana}, C. and {Butler}, R.~C. and {Calabrese}, E. and {Cardoso}, J. -F. and {Catalano}, A. and {Challinor}, A. and {Chamballu}, A. and {Chary}, R. -R. and {Chiang}, H.~C. and {Chluba}, J. and {Christensen}, P.~R. and {Church}, S. and {Clements}, D.~L. and {Colombi}, S. and {Colombo}, L.~P.~L. and {Combet}, C. and {Coulais}, A. and {Crill}, B.~P. and {Curto}, A. and {Cuttaia}, F. and {Danese}, L. and {Davies}, R.~D. and {Davis}, R.~J. and {de Bernardis}, P. and {de Rosa}, A. and {de Zotti}, G. and {Delabrouille}, J. and {D{\'e}sert}, F. -X. and {Di Valentino}, E. and {Dickinson}, C. and {Diego}, J.~M. and {Dolag}, K. and {Dole}, H. and {Donzelli}, S. and {Dor{\'e}}, O. and {Douspis}, M. and {Ducout}, A. and {Dunkley}, J. and {Dupac}, X. and {Efstathiou}, G. and {Elsner}, F. and {En{\ss}lin}, T.~A. and {Eriksen}, H.~K. and {Farhang}, M. and {Fergusson}, J. and {Finelli}, F. and {Forni}, O. and {Frailis}, M. and {Fraisse}, A.~A. and {Franceschi}, E. and {Frejsel}, A. and {Galeotta}, S. and {Galli}, S. and {Ganga}, K. and {Gauthier}, C. and {Gerbino}, M. and {Ghosh}, T. and {Giard}, M. and {Giraud-H{\'e}raud}, Y. and {Giusarma}, E. and {Gjerl{\o}w}, E. and {Gonz{\'a}lez-Nuevo}, J. and {G{\'o}rski}, K.~M. and {Gratton}, S. and {Gregorio}, A. and {Gruppuso}, A. and {Gudmundsson}, J.~E. and {Hamann}, J. and {Hansen}, F.~K. and {Hanson}, D. and {Harrison}, D.~L. and {Helou}, G. and {Henrot-Versill{\'e}}, S. and {Hern{\'a}ndez-Monteagudo}, C. and {Herranz}, D. and {Hildebrandt}, S.~R. and {Hivon}, E. and {Hobson}, M. and {Holmes}, W.~A. and {Hornstrup}, A. and {Hovest}, W. and {Huang}, Z. and {Huffenberger}, K.~M. and {Hurier}, G. and {Jaffe}, A.~H. and {Jaffe}, T.~R. and {Jones}, W.~C. and {Juvela}, M. and {Keih{\"a}nen}, E. and {Keskitalo}, R. and {Kisner}, T.~S. and {Kneissl}, R. and {Knoche}, J. and {Knox}, L. and {Kunz}, M. and {Kurki-Suonio}, H. and {Lagache}, G. and {L{\"a}hteenm{\"a}ki}, A. and {Lamarre}, J. -M. and {Lasenby}, A. and {Lattanzi}, M. and {Lawrence}, C.~R. and {Leahy}, J.~P. and {Leonardi}, R. and {Lesgourgues}, J. and {Levrier}, F. and {Lewis}, A. and {Liguori}, M. and {Lilje}, P.~B. and {Linden-V{\o}rnle}, M. and {L{\'o}pez-Caniego}, M. and {Lubin}, P.~M. and {Mac{\'\i}as-P{\'e}rez}, J.~F. and {Maggio}, G. and {Maino}, D. and {Mandolesi}, N. and {Mangilli}, A. and {Marchini}, A. and {Maris}, M. and {Martin}, P.~G. and {Martinelli}, M. and {Mart{\'\i}nez-Gonz{\'a}lez}, E. and {Masi}, S. and {Matarrese}, S. and {McGehee}, P. and {Meinhold}, P.~R. and {Melchiorri}, A. and {Melin}, J. -B. and {Mendes}, L. and {Mennella}, A. and {Migliaccio}, M. and {Millea}, M. and {Mitra}, S. and {Miville-Desch{\^e}nes}, M. -A. and {Moneti}, A. and {Montier}, L. and {Morgante}, G. and {Mortlock}, D. and {Moss}, A. and {Munshi}, D. and {Murphy}, J.~A. and {Naselsky}, P. and {Nati}, F. and {Natoli}, P. and {Netterfield}, C.~B. and {N{\o}rgaard-Nielsen}, H.~U. and {Noviello}, F. and {Novikov}, D. and {Novikov}, I. and {Oxborrow}, C.~A. and {Paci}, F. and {Pagano}, L. and {Pajot}, F. and {Paladini}, R. and {Paoletti}, D. and {Partridge}, B. and {Pasian}, F. and {Patanchon}, G. and {Pearson}, T.~J. and {Perdereau}, O. and {Perotto}, L. and {Perrotta}, F. and {Pettorino}, V. and {Piacentini}, F. and {Piat}, M. and {Pierpaoli}, E. and {Pietrobon}, D. and {Plaszczynski}, S. and {Pointecouteau}, E. and {Polenta}, G. and {Popa}, L. and {Pratt}, G.~W. and {Pr{\'e}zeau}, G.},
        title = "{Planck 2015 results. XIII. Cosmological parameters}",
      journal = {\aap},
         year = 2016,
        month = sep,
       volume = {594},
          eid = {A13},
        pages = {A13},
          doi = {10.1051/0004-6361/201525830},
archivePrefix = {arXiv},
       eprint = {1502.01589},
 primaryClass = {astro-ph.CO},
       adsurl = {https://ui.adsabs.harvard.edu/abs/2016A&A...594A..13P}
}

@ARTICLE{2017Natur.541...58C,
       author = {{Chatterjee}, S. and {Law}, C.~J. and {Wharton}, R.~S. and {Burke-Spolaor}, S. and {Hessels}, J.~W.~T. and {Bower}, G.~C. and {Cordes}, J.~M. and {Tendulkar}, S.~P. and {Bassa}, C.~G. and {Demorest}, P. and {Butler}, B.~J. and {Seymour}, A. and {Scholz}, P. and {Abruzzo}, M.~W. and {Bogdanov}, S. and {Kaspi}, V.~M. and {Keimpema}, A. and {Lazio}, T.~J.~W. and {Marcote}, B. and {McLaughlin}, M.~A. and {Paragi}, Z. and {Ransom}, S.~M. and {Rupen}, M. and {Spitler}, L.~G. and {van Langevelde}, H.~J.},
        title = "{A direct localization of a fast radio burst and its host}",
      journal = {\nat},
         year = 2017,
        month = jan,
       volume = {541},
       number = {7635},
        pages = {58-61},
          doi = {10.1038/nature20797},
archivePrefix = {arXiv},
       eprint = {1701.01098},
 primaryClass = {astro-ph.HE},
       adsurl = {https://ui.adsabs.harvard.edu/abs/2017Natur.541...58C}
}

@ARTICLE{Ryder2023,
       author = {{Ryder}, S.~D. and {Bannister}, K.~W. and {Bhandari}, S. and {Deller}, A.~T. and {Ekers}, R.~D. and {Glowacki}, M. and {Gordon}, A.~C. and {Gourdji}, K. and {James}, C.~W. and {Kilpatrick}, C.~D. and {Lu}, W. and {Marnoch}, L. and {Moss}, V.~A. and {Prochaska}, J.~X. and {Qiu}, H. and {Sadler}, E.~M. and {Simha}, S. and {Sammons}, M.~W. and {Scott}, D.~R. and {Tejos}, N. and {Shannon}, R.~M.},
        title = "{A luminous fast radio burst that probes the Universe at redshift 1}",
      journal = {Science},
         year = 2023,
        month = oct,
       volume = {382},
       number = {6668},
        pages = {294-299},
          doi = {10.1126/science.adf2678},
archivePrefix = {arXiv},
       eprint = {2210.04680},
 primaryClass = {astro-ph.HE},
       adsurl = {https://ui.adsabs.harvard.edu/abs/2023Sci...382..294R}
}

@article{michilli_extreme_2018,
       author = {{Michilli}, D. and {Seymour}, A. and {Hessels}, J.~W.~T. and {Spitler}, L.~G. and {Gajjar}, V. and {Archibald}, A.~M. and {Bower}, G.~C. and {Chatterjee}, S. and {Cordes}, J.~M. and {Gourdji}, K. and {Heald}, G.~H. and {Kaspi}, V.~M. and {Law}, C.~J. and {Sobey}, C. and {Adams}, E.~A.~K. and {Bassa}, C.~G. and {Bogdanov}, S. and {Brinkman}, C. and {Demorest}, P. and {Fernandez}, F. and {Hellbourg}, G. and {Lazio}, T.~J.~W. and {Lynch}, R.~S. and {Maddox}, N. and {Marcote}, B. and {McLaughlin}, M.~A. and {Paragi}, Z. and {Ransom}, S.~M. and {Scholz}, P. and {Siemion}, A.~P.~V. and {Tendulkar}, S.~P. and {van Rooy}, P. and {Wharton}, R.~S. and {Whitlow}, D.},
        title = "{An extreme magneto-ionic environment associated with the fast radio burst source FRB 121102}",
      journal = {\nat},
         year = 2018,
        month = jan,
       volume = {553},
       number = {7687},
        pages = {182-185},
          doi = {10.1038/nature25149},
archivePrefix = {arXiv},
       eprint = {1801.03965},
 primaryClass = {astro-ph.HE},
       adsurl = {https://ui.adsabs.harvard.edu/abs/2018Natur.553..182M}
}

@ARTICLE{2022Natur.602..585K,
       author = {{Kirsten}, F. and {Marcote}, B. and {Nimmo}, K. and {Hessels}, J.~W.~T. and {Bhardwaj}, M. and {Tendulkar}, S.~P. and {Keimpema}, A. and {Yang}, J. and {Snelders}, M.~P. and {Scholz}, P. and {Pearlman}, A.~B. and {Law}, C.~J. and {Peters}, W.~M. and {Giroletti}, M. and {Paragi}, Z. and {Bassa}, C. and {Hewitt}, D.~M. and {Bach}, U. and {Bezrukovs}, V. and {Burgay}, M. and {Buttaccio}, S.~T. and {Conway}, J.~E. and {Corongiu}, A. and {Feiler}, R. and {Forss{\'e}n}, O. and {Gawro{\'n}ski}, M.~P. and {Karuppusamy}, R. and {Kharinov}, M.~A. and {Lindqvist}, M. and {Maccaferri}, G. and {Melnikov}, A. and {Ould-Boukattine}, O.~S. and {Possenti}, A. and {Surcis}, G. and {Wang}, N. and {Yuan}, J. and {Aggarwal}, K. and {Anna-Thomas}, R. and {Bower}, G.~C. and {Blaauw}, R. and {Burke-Spolaor}, S. and {Cassanelli}, T. and {Clarke}, T.~E. and {Fonseca}, E. and {Gaensler}, B.~M. and {Gopinath}, A. and {Kaspi}, V.~M. and {Kassim}, N. and {Lazio}, T.~J.~W. and {Leung}, C. and {Li}, D.~Z. and {Lin}, H.~H. and {Masui}, K.~W. and {Mckinven}, R. and {Michilli}, D. and {Mikhailov}, A.~G. and {Ng}, C. and {Orbidans}, A. and {Pen}, U.~L. and {Petroff}, E. and {Rahman}, M. and {Ransom}, S.~M. and {Shin}, K. and {Smith}, K.~M. and {Stairs}, I.~H. and {Vlemmings}, W.},
        title = "{A repeating fast radio burst source in a globular cluster}",
      journal = {\nat},
         year = 2022,
        month = feb,
       volume = {602},
       number = {7898},
        pages = {585-589},
          doi = {10.1038/s41586-021-04354-w},
archivePrefix = {arXiv},
       eprint = {2105.11445},
 primaryClass = {astro-ph.HE},
       adsurl = {https://ui.adsabs.harvard.edu/abs/2022Natur.602..585K}
}

@ARTICLE{2003ApJ...594..186B,
       author = {{Blanton}, Michael R. and {Hogg}, David W. and {Bahcall}, Neta A. and {Baldry}, Ivan K. and {Brinkmann}, J. and {Csabai}, Istv{\'a}n and {Eisenstein}, Daniel and {Fukugita}, Masataka and {Gunn}, James E. and {Ivezi{\'c}}, {\v{Z}}eljko and {Lamb}, D.~Q. and {Lupton}, Robert H. and {Loveday}, Jon and {Munn}, Jeffrey A. and {Nichol}, R.~C. and {Okamura}, Sadanori and {Schlegel}, David J. and {Shimasaku}, Kazuhiro and {Strauss}, Michael A. and {Vogeley}, Michael S. and {Weinberg}, David H.},
        title = "{The Broadband Optical Properties of Galaxies with Redshifts 0.02<z<0.22}",
      journal = {\apj},
         year = 2003,
        month = sep,
       volume = {594},
       number = {1},
        pages = {186-207},
          doi = {10.1086/375528},
archivePrefix = {arXiv},
       eprint = {astro-ph/0209479},
 primaryClass = {astro-ph},
       adsurl = {https://ui.adsabs.harvard.edu/abs/2003ApJ...594..186B}
}

@ARTICLE{2023ApJ...954...80G,
       author = {{Gordon}, Alexa C. and {Fong}, Wen-fai and {Kilpatrick}, Charles D. and {Eftekhari}, Tarraneh and {Leja}, Joel and {Prochaska}, J. Xavier and {Nugent}, Anya E. and {Bhandari}, Shivani and {Blanchard}, Peter K. and {Caleb}, Manisha and {Day}, Cherie K. and {Deller}, Adam T. and {Dong}, Yuxin and {Glowacki}, Marcin and {Gourdji}, Kelly and {Mannings}, Alexandra G. and {Mahoney}, Elizabeth K. and {Marnoch}, Lachlan and {Miller}, Adam A. and {Paterson}, Kerry and {Rastinejad}, Jillian C. and {Ryder}, Stuart D. and {Sadler}, Elaine M. and {Scott}, Danica R. and {Sears}, Huei and {Shannon}, Ryan M. and {Simha}, Sunil and {Stappers}, Benjamin W. and {Tejos}, Nicolas},
        title = "{The Demographics, Stellar Populations, and Star Formation Histories of Fast Radio Burst Host Galaxies: Implications for the Progenitors}",
      journal = {\apj},
         year = 2023,
        month = sep,
       volume = {954},
       number = {1},
          eid = {80},
        pages = {80},
          doi = {10.3847/1538-4357/ace5aa},
archivePrefix = {arXiv},
       eprint = {2302.05465},
 primaryClass = {astro-ph.GA},
       adsurl = {https://ui.adsabs.harvard.edu/abs/2023ApJ...954...80G}
}

@ARTICLE{ne2025,
       author = {{Ocker}, Stella Koch and {Cordes}, James M.},
        title = "{NE2025: An Updated Electron Density Model for the Galactic Interstellar Medium}",
      journal = {\apj},
         year = 2026,
        month = may,
       volume = {1002},
       number = {1},
          eid = {3},
        pages = {3},
          doi = {10.3847/1538-4357/ae5825},
archivePrefix = {arXiv},
       eprint = {2602.11838},
 primaryClass = {astro-ph.GA},
       adsurl = {https://ui.adsabs.harvard.edu/abs/2026ApJ..1002....3O}
}

@ARTICLE{2026arXiv260501994S,
       author = {{Sharma}, Kritti and {Ravi}, Vikram and {Anbajagane}, Dhayaa and {Coulton}, William R. and {Krause}, Elisabeth and {Schuster}, Nico and {Pisani}, Alice and {McCarty}, Samuel and {Connor}, Liam and {Ferraro}, Simone and {Hamaus}, Nico and {S}, Pranjal R.},
        title = "{Baryons in the Darkest Sites of the Universe}",
      journal = {arXiv e-prints},
         year = 2026,
        month = may,
          eid = {arXiv:2605.01994},
        pages = {arXiv:2605.01994},
          doi = {10.48550/arXiv.2605.01994},
archivePrefix = {arXiv},
       eprint = {2605.01994},
 primaryClass = {astro-ph.CO},
       adsurl = {https://ui.adsabs.harvard.edu/abs/2026arXiv260501994S}
}

@ARTICLE{2026ApJ...996...78H,
       author = {{Horowicz}, Asaf and {Margalit}, Ben},
        title = "{The Host Galaxies of Fast Radio Bursts Track a Combination of Stellar Mass and Star Formation, Similar to Type Ia Supernovae}",
      journal = {\apj},
         year = 2026,
        month = jan,
       volume = {996},
       number = {1},
          eid = {78},
        pages = {78},
          doi = {10.3847/1538-4357/ae1f11},
archivePrefix = {arXiv},
       eprint = {2504.08038},
 primaryClass = {astro-ph.HE},
       adsurl = {https://ui.adsabs.harvard.edu/abs/2026ApJ...996...78H}
}

@ARTICLE{PATH,
       author = {{Aggarwal}, Kshitij and {Budav{\'a}ri}, Tam{\'a}s and {Deller}, Adam T. and {Eftekhari}, Tarraneh and {James}, Clancy W. and {Prochaska}, J. Xavier and {Tendulkar}, Shriharsh P.},
        title = "{Probabilistic Association of Transients to their Hosts (PATH)}",
      journal = {\apj},
         year = 2021,
        month = apr,
       volume = {911},
       number = {2},
          eid = {95},
        pages = {95},
          doi = {10.3847/1538-4357/abe8d2},
archivePrefix = {arXiv},
       eprint = {2102.10627},
 primaryClass = {astro-ph.HE},
       adsurl = {https://ui.adsabs.harvard.edu/abs/2021ApJ...911...95A}
}

@ARTICLE{2025PASA...42...17H,
       author = {{Hoffmann}, Jordan Luke and {James}, Clancy and {Glowacki}, Marcin and {Prochaska}, Xavier and {Gordon}, Alexa and {Deller}, Adam and {Shannon}, Ryan M. and {Ryder}, Stuart},
        title = "{Modelling DSA, FAST, and CRAFT surveys in a z-DM analysis and constraining a minimum FRB energy}",
      journal = {\pasa},
         year = 2025,
        month = jan,
       volume = {42},
          eid = {e017},
        pages = {e017},
          doi = {10.1017/pasa.2024.127},
archivePrefix = {arXiv},
       eprint = {2408.04878},
 primaryClass = {astro-ph.CO},
       adsurl = {https://ui.adsabs.harvard.edu/abs/2025PASA...42...17H}
}

@ARTICLE{HoffmannHalo26,
       author = {{Hoffmann}, Jordan Luke and {James}, Clancy and {Prochaska}, Xavier and {Glowacki}, Marcin},
        title = "{I can see your halo: Constraining the MilkyWay halo DM with FRB population studies}",
      journal = {\pasa},
         year = 2026,
        month = jan,
       volume = {43},
          eid = {e017},
        pages = {e017},
          doi = {10.1017/pasa.2026.10143},
archivePrefix = {arXiv},
       eprint = {2601.05496},
 primaryClass = {astro-ph.GA},
       adsurl = {https://ui.adsabs.harvard.edu/abs/2026PASA...43...17H}
}

@ARTICLE{2025arXiv250801648C,
       author = {{Caleb}, Manisha and {Nanayakkara}, Themiya and {Stappers}, Benjamin and {Pastor-Marazuela}, In{\'e}s and {Khrykin}, Ilya S. and {Glazebrook}, Karl and {Tejos}, Nicolas and {Prochaska}, J. Xavier and {Rajwade}, Kaustubh and {Mas-Ribas}, Lluis and {Driessen}, Laura N. and {Fong}, Wen-fai and {Gordon}, Alexa C. and {Hoffmann}, Jordan and {James}, Clancy W. and {Jankowski}, Fabian and {Kahinga}, Lordrick and {Kramer}, Michael and {Simha}, Sunil and {Barr}, Ewan D. and {Christiaan Bezuidenhout}, Mechiel and {Deng}, Xihan and {Lin}, Zeren and {Marnoch}, Lachlan and {Martin}, Christopher D. and {Nugent}, Anya and {Shaji}, Kavya and {Tian}, Jun},
        title = "{A fast radio burst from the first 3 billion years of the Universe}",
      journal = {arXiv e-prints},
         year = 2025,
        month = aug,
          eid = {arXiv:2508.01648},
        pages = {arXiv:2508.01648},
          doi = {10.48550/arXiv.2508.01648},
archivePrefix = {arXiv},
       eprint = {2508.01648},
 primaryClass = {astro-ph.HE},
       adsurl = {https://ui.adsabs.harvard.edu/abs/2025arXiv250801648C}
}

@ARTICLE{2020MNRAS.494..665L,
       author = {{Luo}, Rui and {Men}, Yunpeng and {Lee}, Kejia and {Wang}, Weiyang and {Lorimer}, D.~R. and {Zhang}, Bing},
        title = "{On the FRB luminosity function - - II. Event rate density}",
      journal = {\mnras},
         year = 2020,
        month = may,
       volume = {494},
       number = {1},
        pages = {665-679},
          doi = {10.1093/mnras/staa704},
archivePrefix = {arXiv},
       eprint = {2003.04848},
 primaryClass = {astro-ph.HE},
       adsurl = {https://ui.adsabs.harvard.edu/abs/2020MNRAS.494..665L}
}

@ARTICLE{2025ApJS..280....6C,
       author = {{CHIME/FRB Collaboration} and {Amiri}, Mandana and {Amouyal}, Daniel and {Andersen}, Bridget C. and {Andrew}, Shion and {Bandura}, Kevin and {Bhardwaj}, Mohit and {Boyle}, P.~J. and {Brar}, Charanjot and {Cassity}, Alyssa and {Chatterjee}, Shami and others},
        title = "{A Catalog of Local Universe Fast Radio Bursts from CHIME/FRB and the KKO}",
      journal = {\apjs},
         year = 2025,
        month = sep,
       volume = {280},
       number = {1},
          eid = {6},
        pages = {6},
          doi = {10.3847/1538-4365/addbda},
archivePrefix = {arXiv},
       eprint = {2502.11217},
 primaryClass = {astro-ph.HE},
       adsurl = {https://ui.adsabs.harvard.edu/abs/2025ApJS..280....6C}
}

@ARTICLE{DSA_Sharma_sfr,
       author = {{Sharma}, Kritti and {Ravi}, Vikram and {Connor}, Liam and {Law}, Casey and {Ocker}, Stella Koch and {Sherman}, Myles and {Kosogorov}, Nikita and {Faber}, Jakob and {Hallinan}, Gregg and {Harnach}, Charlie and {Hellbourg}, Greg and {Hobbs}, Rick and {Hodge}, David and {Hodges}, Mark and {Lamb}, James and {Rasmussen}, Paul and {Somalwar}, Jean and {Weinreb}, Sander and {Woody}, David and {Leja}, Joel and {Anand}, Shreya and {Das}, Kaustav Kashyap and {Qin}, Yu-Jing and {Rose}, Sam and {Dong}, Dillon Z. and {Miller}, Jessie and {Yao}, Yuhan},
        title = "{Preferential occurrence of fast radio bursts in massive star-forming galaxies}",
      journal = {\nat},
         year = 2024,
        month = nov,
       volume = {635},
       number = {8037},
        pages = {61-66},
          doi = {10.1038/s41586-024-08074-9},
archivePrefix = {arXiv},
       eprint = {2409.16964},
 primaryClass = {astro-ph.HE},
       adsurl = {https://ui.adsabs.harvard.edu/abs/2024Natur.635...61S}
}

@ARTICLE{Shannon_FE,
       author = {{Shannon}, R.~M. and {Macquart}, J.-P. and {Bannister}, K.~W. and {Ekers}, R.~D. and {James}, C.~W. and {Os{\l}owski}, S. and {Qiu}, H. and {Sammons}, M. and {Hotan}, A.~W. and {Voronkov}, M.~A. and {Beresford}, R.~J. and {Brothers}, M. and {Brown}, A.~J. and {Bunton}, J.~D. and {Chippendale}, A.~P. and {Haskins}, C. and {Leach}, M. and {Marquarding}, M. and {McConnell}, D. and {Pilawa}, M.~A. and {Sadler}, E.~M. and {Troup}, E.~R. and {Tuthill}, J. and {Whiting}, M.~T. and {Allison}, J.~R. and {Anderson}, C.~S. and {Bell}, M.~E. and {Collier}, J.~D. and {G{\"u}rkan}, G. and {Heald}, G. and {Riseley}, C.~J.},
        title = "{The dispersion-brightness relation for fast radio bursts from a wide-field survey}",
      journal = {\nat},
         year = 2018,
        month = oct,
       volume = {562},
       number = {7727},
        pages = {386-390},
          doi = {10.1038/s41586-018-0588-y},
       adsurl = {https://ui.adsabs.harvard.edu/abs/2018Natur.562..386S}
}

@ARTICLE{Shannon_ICS,
       author = {{Shannon}, Ryan M. and {Bannister}, Keith W. and {Bera}, Apurba and {Bhandari}, Shivani and {Day}, Cherie K. and {Deller}, Adam T. and {Dial}, Tyson and {Dobie}, Dougal and {Ekers}, Ron D. and {Fong}, Wen-fai and {Glowacki}, Marcin and {Gordon}, Alexa C. and {Gourdji}, Kelly and {Jaini}, Akhil and {James}, Clancy W. and {Kumar}, Pravir and {Mahony}, Elizabeth K. and {Marnoch}, Lachlan and {Muller}, August R. and {Prochaska}, Xavier and {Qiu}, Hao and {Ryder}, Stuart D. and {Sadler}, Elaine M. and {Scott}, Danica R. and {Tejos}, N. and {Uttarkar}, Pavan A. and {Wang}, Yuanming},
        title = "{The commensal real-time ASKAP fast transient incoherent-sum survey}",
      journal = {\pasa},
         year = 2025,
        month = jan,
       volume = {42},
          eid = {e036},
        pages = {e036},
          doi = {10.1017/pasa.2025.8},
       adsurl = {https://ui.adsabs.harvard.edu/abs/2025PASA...42...36S}
}

@article{james_zdm_2022,
	title = {The z–{DM} distribution of fast radio bursts},
	volume = {509},
	issn = {0035-8711},
	url = {https://doi.org/10.1093/mnras/stab3051},
	doi = {10.1093/mnras/stab3051},
	number = {4},
	urldate = {2023-08-31},
	journal = {\mnras},
	author = {James, C. W. and Prochaska, J. X. and Macquart, J.-P. and North-Hickey, F. O. and Bannister, K. W. and Dunning, A.},
	month = feb,
	year = {2022},
	pages = {4775--4802},
}

@ARTICLE{2024ApJ...963L..34G,
       author = {{Gordon}, Alexa C. and {Fong}, Wen-fai and {Simha}, Sunil and {Dong}, Yuxin and {Kilpatrick}, Charles D. and {Deller}, Adam T. and {Ryder}, Stuart D. and {Eftekhari}, Tarraneh and {Glowacki}, Marcin and {Marnoch}, Lachlan and {Muller}, August R. and {Nugent}, Anya E. and {Palmese}, Antonella and {Prochaska}, J. Xavier and {Rafelski}, Marc and {Shannon}, Ryan M. and {Tejos}, Nicolas},
        title = "{A Fast Radio Burst in a Compact Galaxy Group at z {\ensuremath{\sim}} 1}",
      journal = {\apjl},
         year = 2024,
        month = mar,
       volume = {963},
       number = {2},
          eid = {L34},
        pages = {L34},
          doi = {10.3847/2041-8213/ad2773},
archivePrefix = {arXiv},
       eprint = {2311.10815},
 primaryClass = {astro-ph.GA},
       adsurl = {https://ui.adsabs.harvard.edu/abs/2024ApJ...963L..34G}
}

@ARTICLE{2026MNRAS.545f2144P,
       author = {{Pastor-Marazuela}, In{\'e}s and {Gordon}, Alexa C. and {Stappers}, Ben and {Khrykin}, Ilya S. and {Tejos}, Nicolas and {Rajwade}, Kaustubh and {Caleb}, Manisha and {Surnis}, Mayuresh P. and {Driessen}, Laura N. and {Simha}, Sunil and {Tian}, Jun and {Prochaska}, J. Xavier and {Barr}, Ewan and {Buchner}, Sarah and {Fong}, Wen-Fai and {Jankowski}, Fabian and {Kahinga}, Lordrick and {Kilpatrick}, Charles D. and {Kramer}, Michael and {Mas-Ribas}, Lluis and {Hennawi}, Joseph},
        title = "{Localization and host galaxy identification of new fast radio bursts with MeerKAT}",
      journal = {\mnras},
         year = 2026,
        month = feb,
       volume = {545},
       number = {4},
          eid = {staf2144},
        pages = {staf2144},
          doi = {10.1093/mnras/staf2144},
archivePrefix = {arXiv},
       eprint = {2507.05982},
 primaryClass = {astro-ph.HE},
       adsurl = {https://ui.adsabs.harvard.edu/abs/2026MNRAS.545f2144P}
}

@ARTICLE{2025PASA...42..157G,
       author = {{Glowacki}, Marcin and {Bera}, Apurba and {James}, Clancy and {Patterson}, Jasper and {Deller}, Adam T. and {Gordon}, Alexa and {Marnoch}, Lachlan and {Muller}, August and {Prochaska}, Xavier and {Ryder}, Stuart and {Shannon}, Ryan M. and {Tejos}, Nicolas and {Mannings}, Alexandra G.},
        title = "{An investigation into correlations between FRB and host galaxy properties}",
      journal = {\pasa},
         year = 2025,
        month = nov,
       volume = {42},
          eid = {e157},
        pages = {e157},
          doi = {10.1017/pasa.2025.10122},
archivePrefix = {arXiv},
       eprint = {2506.23403},
 primaryClass = {astro-ph.HE},
       adsurl = {https://ui.adsabs.harvard.edu/abs/2025PASA...42..157G}
}

@ARTICLE{2026ApJ...997L...5P,
       author = {{Patil}, Swarali Shivraj and {Main}, Robert A. and {Fonseca}, Emmanuel and {McGregor}, Kyle and {Gaensler}, B.~M. and {Bhardwaj}, Mohit and {Brar}, Charanjot and {Cook}, Amanda M. and {Curtin}, Alice P. and {Eadie}, Gwendolyn and {Joseph}, Ronniy and {Kahinga}, Lordrick and {Kaspi}, Victoria and {Khan}, Afrokk and {Kharel}, Bikash and {Lanman}, Adam E. and {Leung}, Calvin and {Masui}, Kiyoshi W. and {Ng}, Mason and {Nimmo}, Kenzie and {Pandhi}, Ayush and {Pearlman}, Aaron B. and {Pleunis}, Ziggy and {Sammons}, Mawson W. and {Sand}, Ketan R. and {Scholz}, Paul and {Shin}, Kaitlyn and {Siegel}, Seth R. and {Smith}, Kendrick},
        title = "{A Spatial Gap in the Sky Distribution of Fast Radio Burst Detections Coinciding with Galactic Plasma Overdensities}",
      journal = {\apjl},
         year = 2026,
        month = jan,
       volume = {997},
       number = {1},
          eid = {L5},
        pages = {L5},
          doi = {10.3847/2041-8213/ae2eb3},
archivePrefix = {arXiv},
       eprint = {2509.06721},
 primaryClass = {astro-ph.HE},
       adsurl = {https://ui.adsabs.harvard.edu/abs/2026ApJ...997L...5P}
}

@ARTICLE{CHIMECat2,
       author = {{CHIME/FRB Collaboration} and {Abbott}, Thomas and {Andersen}, Bridget C. and {Andrew}, Shion and {Bandura}, Kevin and {Bhardwaj}, Mohit and {Bhusare}, Yash and {Brar}, Charanjot and {Cassanelli}, Tomas and {Chatterjee}, Shami and {Cliche}, Jean-Francois and {Cook}, Amanda M. and {Curtin}, Alice and {Dobbs}, Matt and {Dong}, Fengqiu Adam and {Eadie}, Gwendolyn and {Eftekhari}, Tarraneh and {Fonseca}, Emmanuel and {Gaensler}, B.~M. and {Good}, Deborah and {Halpern}, Mark and {Hessels}, Jason W.~T. and {Ibik}, Adaeze and {Jain}, Naman and {Joseph}, Ronniy C. and {Kader}, Zarif and {Kaspi}, Victoria M. and {Khan}, Afrokk and {Kharel}, Bikash and {Kumar}, Ajay and {Landecker}, T.~L. and {Lang}, Dustin and {Lanman}, Adam E. and {L'Argent}, Magnus and {Lazda}, Mattias and {Leung}, Calvin and {Li}, Dong Zi and {Lintott}, Chris J. and {Main}, Robert and {Masui}, Kiyoshi W. and {Mate}, Sujay and {McGregor}, Kyle and {Mckinven}, Ryan and {Mena-Parra}, Juan and {Meyers}, Bradley W. and {Michilli}, Daniele and {Ng}, Cherry and {Ng}, Mason and {Nimmo}, Kenzie and {Noble}, Gavin and {Pandhi}, Ayush and {Patil}, Swarali S. and {Pearlman}, Aaron B. and {Pen}, Ue-Li and {Pleunis}, Ziggy and {Prochaska}, J. Xavier and {Rafiei-Ravandi}, Masoud and {Ransom}, Scott and {Renard}, Andre and {Sammons}, Mawson W. and {Sand}, Ketan R. and {Scholz}, Paul and {Shah}, Vishwangi and {Shin}, Kaitlyn and {Siegel}, Seth R. and {Sirota}, Sloane and {Smith}, Kendrick and {Stairs}, Ingrid and {Stenning}, David C. and {Tendulkar}, Shriharsh P. and {Vanderlinde}, Keith and {Walmsley}, Mike and {Wang}, Haochen and {Wulf}, Dallas},
        title = "{The Second CHIME/FRB Catalog of Fast Radio Bursts}",
      journal = {arXiv e-prints},
         year = 2026,
        month = jan,
          eid = {arXiv:2601.09399},
        pages = {arXiv:2601.09399},
          doi = {10.48550/arXiv.2601.09399},
archivePrefix = {arXiv},
       eprint = {2601.09399},
 primaryClass = {astro-ph.HE},
       adsurl = {https://ui.adsabs.harvard.edu/abs/2026arXiv260109399F}
}

@ARTICLE{2024ApJ...971L..51B,
       author = {{Bhardwaj}, Mohit and {Michilli}, Daniele and {Kirichenko}, Aida Yu. and {Modilim}, Obinna and {Shin}, Kaitlyn and {Kaspi}, Victoria M. and {Andersen}, Bridget C. and {Cassanelli}, Tomas and {Brar}, Charanjot and {Chatterjee}, Shami and {Cook}, Amanda M. and {Dong}, Fengqiu Adam and {Fonseca}, Emmanuel and {Gaensler}, B.~M. and {Ibik}, Adaeze L. and {Kaczmarek}, J.~F. and {Lanman}, Adam E. and {Leung}, Calvin and {Masui}, K.~W. and {Pandhi}, Ayush and {Pearlman}, Aaron B. and {Petroff}, Emily and {Pleunis}, Ziggy and {Prochaska}, J. Xavier and {Rafiei-Ravandi}, Masoud and {Sand}, Ketan R. and {Scholz}, Paul and {Smith}, Kendrick M.},
        title = "{Host Galaxies for Four Nearby CHIME/FRB Sources and the Local Universe FRB Host Galaxy Population}",
      journal = {\apjl},
         year = 2024,
        month = aug,
       volume = {971},
       number = {2},
          eid = {L51},
        pages = {L51},
          doi = {10.3847/2041-8213/ad64d1},
archivePrefix = {arXiv},
       eprint = {2310.10018},
 primaryClass = {astro-ph.HE},
       adsurl = {https://ui.adsabs.harvard.edu/abs/2024ApJ...971L..51B}
}

@ARTICLE{Marnoch2023,
       author = {{Marnoch}, Lachlan and {Ryder}, Stuart D. and {James}, Clancy W. and {Gordon}, Alexa C. and {Sammons}, Mawson W. and {Prochaska}, J. Xavier and {Tejos}, Nicolas and {Deller}, Adam T. and {Scott}, Danica R. and {Bhandari}, Shivani and {Glowacki}, Marcin and {Mahony}, Elizabeth K. and {McDermid}, Richard M. and {Sadler}, Elaine M. and {Shannon}, Ryan M. and {Qiu}, Hao},
        title = "{The unseen host galaxy and high dispersion measure of a precisely localized fast radio burst suggests a high-redshift origin}",
      journal = {\mnras},
         year = 2023,
        month = oct,
       volume = {525},
       number = {1},
        pages = {994-1007},
          doi = {10.1093/mnras/stad2353},
archivePrefix = {arXiv},
       eprint = {2307.14702},
 primaryClass = {astro-ph.HE},
       adsurl = {https://ui.adsabs.harvard.edu/abs/2023MNRAS.525..994M}
}

@ARTICLE{2017ApJ...837..170L,
       author = {{Leja}, Joel and {Johnson}, Benjamin D. and {Conroy}, Charlie and {van Dokkum}, Pieter G. and {Byler}, Nell},
        title = "{Deriving Physical Properties from Broadband Photometry with Prospector: Description of the Model and a Demonstration of its Accuracy Using 129 Galaxies in the Local Universe}",
      journal = {\apj},
         year = 2017,
        month = mar,
       volume = {837},
       number = {2},
          eid = {170},
        pages = {170},
          doi = {10.3847/1538-4357/aa5ffe},
archivePrefix = {arXiv},
       eprint = {1609.09073},
 primaryClass = {astro-ph.GA},
       adsurl = {https://ui.adsabs.harvard.edu/abs/2017ApJ...837..170L}
}

@ARTICLE{Cordes_McLaughlin_2003,
   author = {{Cordes}, J.~M. and {McLaughlin}, M.~A.},
    title = "{Searches for Fast Radio Transients}",
  journal = {\apj},
   eprint = {astro-ph/0304364},
     year = 2003,
    month = oct,
   volume = 596,
    pages = {1142-1154},
      doi = {10.1086/378231},
   adsurl = {https://ui.adsabs.harvard.edu/abs/2003ApJ...596.1142C}
}

@ARTICLE{Macquart2020,
       author = {{Macquart}, J.-P. and {Prochaska}, J.~X. and {McQuinn}, M. and
         {Bannister}, K.~W. and {Bhandari}, S. and {Day}, C.~K. and
         {Deller}, A.~T. and {Ekers}, R.~D. and {James}, C.~W. and
         {Marnoch}, L. and {Os{\l}owski}, S. and {Phillips}, C. and
         {Ryder}, S.~D. and {Scott}, D.~R. and {Shannon}, R.~M. and {Tejos}, N.},
        title = "{A census of baryons in the Universe from localized fast radio bursts}",
      journal = {\nat},
         year = 2020,
        month = may,
       volume = {581},
       number = {7809},
        pages = {391-395},
          doi = {10.1038/s41586-020-2300-2},
archivePrefix = {arXiv},
       eprint = {2005.13161},
 primaryClass = {astro-ph.CO},
       adsurl = {https://ui.adsabs.harvard.edu/abs/2020Natur.581..391M}
}

@ARTICLE{2026arXiv260300371J,
       author = {{James}, C.~W. and {Smith}, B. and {Dage}, K. and {Chies Santos}, A.~L. and {Bannister}, K.~W. and {Caleb}, M. and {Crenshaw}, J.~F. and {Deller}, A.~T. and {Lee}, K.~G. and {Marnoch}, L. and {Rajwade}, K.~M. and {Ryder}, S.~D. and {Shannon}, R.~M. and {Stappers}, B. and {Zhang}, T.},
        title = "{Fast Radio Bursts in the Era of the Vera C. Rubin Observatory's Legacy Survey of Space and Time}",
      journal = {arXiv e-prints},
         year = 2026,
        month = feb,
          eid = {arXiv:2603.00371},
        pages = {arXiv:2603.00371},
archivePrefix = {arXiv},
       eprint = {2603.00371},
 primaryClass = {astro-ph.HE},
       adsurl = {https://ui.adsabs.harvard.edu/abs/2026arXiv260300371J}
}

@ARTICLE{2024Natur.634.1065B,
       author = {{Bhardwaj}, Mohit and {Lee}, Jimin and {Ji}, Kevin},
        title = "{Selection bias obfuscates the discovery of fast radio burst sources}",
      journal = {\nat},
         year = 2024,
        month = oct,
       volume = {634},
       number = {8036},
        pages = {1065-1069},
          doi = {10.1038/s41586-024-08065-w},
archivePrefix = {arXiv},
       eprint = {2408.01876},
 primaryClass = {astro-ph.HE},
       adsurl = {https://ui.adsabs.harvard.edu/abs/2024Natur.634.1065B}
}

@ARTICLE{2023ApJ...944..105S,
       author = {{Shin}, Kaitlyn and {Masui}, Kiyoshi W. and {Bhardwaj}, Mohit and {Cassanelli}, Tomas and {Chawla}, Pragya and {Dobbs}, Matt and {Dong}, Fengqiu Adam and {Fonseca}, Emmanuel and {Gaensler}, B.~M. and {Herrera-Mart{\'\i}n}, Antonio and {Kaczmarek}, Jane and {Kaspi}, Victoria and {Leung}, Calvin and {Merryfield}, Marcus and {Michilli}, Daniele and {M{\"u}nchmeyer}, Moritz and {Pearlman}, Aaron B. and {Rafiei-Ravandi}, Masoud and {Smith}, Kendrick and {Stairs}, Ingrid and {Tendulkar}, Shriharsh P.},
        title = "{Inferring the Energy and Distance Distributions of Fast Radio Bursts Using the First CHIME/FRB Catalog}",
      journal = {\apj},
         year = 2023,
        month = feb,
       volume = {944},
       number = {1},
          eid = {105},
        pages = {105},
          doi = {10.3847/1538-4357/acaf06},
archivePrefix = {arXiv},
       eprint = {2207.14316},
 primaryClass = {astro-ph.HE},
       adsurl = {https://ui.adsabs.harvard.edu/abs/2023ApJ...944..105S}
}

@ARTICLE{FRBPOPPY,
       author = {{Gardenier}, D.~W. and {van Leeuwen}, J. and {Connor}, L. and {Petroff}, E.},
        title = "{Synthesising the intrinsic FRB population using frbpoppy}",
      journal = {\aap},
         year = 2019,
        month = dec,
       volume = {632},
          eid = {A125},
        pages = {A125},
          doi = {10.1051/0004-6361/201936404},
archivePrefix = {arXiv},
       eprint = {1910.08365},
 primaryClass = {astro-ph.HE},
       adsurl = {https://ui.adsabs.harvard.edu/abs/2019A&A...632A.125G}
}

@ARTICLE{Bhandari+22,
       author = {{Bhandari}, Shivani and {Heintz}, Kasper E. and {Aggarwal}, Kshitij and {Marnoch}, Lachlan and {Day}, Cherie K. and {Sydnor}, Jessica and {Burke-Spolaor}, Sarah and {Law}, Casey J. and {Xavier Prochaska}, J. and {Tejos}, Nicolas and {Bannister}, Keith W. and {Butler}, Bryan J. and {Deller}, Adam T. and {Ekers}, R.~D. and {Flynn}, Chris and {Fong}, Wen-fai and {James}, Clancy W. and {Lazio}, T. Joseph W. and {Luo}, Rui and {Mahony}, Elizabeth K. and {Ryder}, Stuart D. and {Sadler}, Elaine M. and {Shannon}, Ryan M. and {Han}, JinLin and {Lee}, Kejia and {Zhang}, Bing},
        title = "{Characterizing the Fast Radio Burst Host Galaxy Population and its Connection to Transients in the Local and Extragalactic Universe}",
      journal = {\aj},
         year = 2022,
        month = feb,
       volume = {163},
       number = {2},
          eid = {69},
        pages = {69},
          doi = {10.3847/1538-3881/ac3aec},
archivePrefix = {arXiv},
       eprint = {2108.01282},
 primaryClass = {astro-ph.HE},
       adsurl = {https://ui.adsabs.harvard.edu/abs/2022AJ....163...69B}
}

@ARTICLE{2024ApJ...962L..13G,
       author = {{Glowacki}, M. and {Bera}, A. and {Lee-Waddell}, K. and {Deller}, A.~T. and {Dial}, T. and {Gourdji}, K. and {Simha}, S. and {Caleb}, M. and {Marnoch}, L. and {Prochaska}, J. Xavier and {Ryder}, S.~D. and {Shannon}, R.~M. and {Tejos}, N.},
        title = "{H I, FRB, What's Your z: The First FRB Host Galaxy Redshift from Radio Observations}",
      journal = {\apjl},
         year = 2024,
        month = feb,
       volume = {962},
       number = {1},
          eid = {L13},
        pages = {L13},
          doi = {10.3847/2041-8213/ad1f62},
archivePrefix = {arXiv},
       eprint = {2311.16808},
 primaryClass = {astro-ph.GA},
       adsurl = {https://ui.adsabs.harvard.edu/abs/2024ApJ...962L..13G}
}

@ARTICLE{Bhandari2020,
       author = {{Bhandari}, Shivani and {Sadler}, Elaine M. and {Prochaska}, J. Xavier and {Simha}, Sunil and {Ryder}, Stuart D. and {Marnoch}, Lachlan and {Bannister}, Keith W. and {Macquart}, Jean-Pierre and {Flynn}, Chris and {Shannon}, Ryan M. and {Tejos}, Nicolas and {Corro-Guerra}, Felipe and {Day}, Cherie K. and {Deller}, Adam T. and {Ekers}, Ron and {Lopez}, Sebastian and {Mahony}, Elizabeth K. and {Nu{\~n}ez}, Consuelo and {Phillips}, Chris},
        title = "{The Host Galaxies and Progenitors of Fast Radio Bursts Localized with the Australian Square Kilometre Array Pathfinder}",
      journal = {\apjl},
         year = 2020,
        month = jun,
       volume = {895},
       number = {2},
          eid = {L37},
        pages = {L37},
          doi = {10.3847/2041-8213/ab672e},
archivePrefix = {arXiv},
       eprint = {2005.13160},
 primaryClass = {astro-ph.GA},
       adsurl = {https://ui.adsabs.harvard.edu/abs/2020ApJ...895L..37B}
}

@ARTICLE{2023MNRAS.524.4275J,
       author = {{Jankowski}, F. and {Bezuidenhout}, M.~C. and {Caleb}, M. and {Driessen}, L.~N. and {Malenta}, M. and {Morello}, V. and {Rajwade}, K.~M. and {Sanidas}, S. and {Stappers}, B.~W. and {Surnis}, M.~P. and {Barr}, E.~D. and {Chen}, W. and {Kramer}, M. and {Wu}, J. and {Buchner}, S. and {Serylak}, M. and {Prochaska}, J. Xavier},
        title = "{A sample of fast radio bursts discovered and localized with MeerTRAP at the MeerKAT telescope}",
      journal = {\mnras},
         year = 2023,
        month = sep,
       volume = {524},
       number = {3},
        pages = {4275-4295},
          doi = {10.1093/mnras/stad2041},
archivePrefix = {arXiv},
       eprint = {2302.10107},
 primaryClass = {astro-ph.HE},
       adsurl = {https://ui.adsabs.harvard.edu/abs/2023MNRAS.524.4275J}
}

@ARTICLE{Farah2018,
       author = {{Farah}, W. and {Flynn}, C. and {Bailes}, M. and {Jameson}, A. and {Bannister}, K.~W. and {Barr}, E.~D. and {Bateman}, T. and {Bhandari}, S. and {Caleb}, M. and {Campbell-Wilson}, D. and {Chang}, S. -W. and {Deller}, A. and {Green}, A.~J. and {Hunstead}, R. and {Jankowski}, F. and {Keane}, E. and {Macquart}, J. -P. and {M{\"o}ller}, A. and {Onken}, C.~A. and {Os{\l}owski}, S. and {Parthasarathy}, A. and {Plant}, K. and {Ravi}, V. and {Shannon}, R.~M. and {Tucker}, B.~E. and {Venkatraman Krishnan}, V. and {Wolf}, C.},
        title = "{FRB microstructure revealed by the real-time detection of FRB170827}",
      journal = {\mnras},
         year = 2018,
        month = jul,
       volume = {478},
       number = {1},
        pages = {1209-1217},
          doi = {10.1093/mnras/sty1122},
archivePrefix = {arXiv},
       eprint = {1803.05697},
 primaryClass = {astro-ph.HE},
       adsurl = {https://ui.adsabs.harvard.edu/abs/2018MNRAS.478.1209F}
}

@ARTICLE{FORS2,
       author = {{Appenzeller}, I. and {Rupprecht}, G.},
        title = "{FORS - the focal reducer for the VLT.}",
      journal = {The Messenger},
         year = 1992,
        month = mar,
       volume = {67},
        pages = {18-21},
       adsurl = {https://ui.adsabs.harvard.edu/abs/1992Msngr..67...18A}
}

@ARTICLE{Loudas25,
       author = {{Loudas}, Nick and {Li}, Dongzi and {Strauss}, Michael A. and {Leja}, Joel},
        title = "{Unveiling the Origin of Fast Radio Bursts by Modeling the Stellar Mass and Star Formation Distributions of Their Host Galaxies}",
      journal = {\apj},
         year = 2025,
        month = sep,
       volume = {991},
       number = {1},
          eid = {85},
        pages = {85},
          doi = {10.3847/1538-4357/adf555},
archivePrefix = {arXiv},
       eprint = {2502.15566},
 primaryClass = {astro-ph.HE},
       adsurl = {https://ui.adsabs.harvard.edu/abs/2025ApJ...991...85L}
}

@ARTICLE{2022ApJ...936..165L,
       author = {{Leja}, Joel and {Speagle}, Joshua S. and {Ting}, Yuan-Sen and {Johnson}, Benjamin D. and {Conroy}, Charlie and {Whitaker}, Katherine E. and {Nelson}, Erica J. and {van Dokkum}, Pieter and {Franx}, Marijn},
        title = "{A New Census of the 0.2 < z < 3.0 Universe. II. The Star-forming Sequence}",
      journal = {\apj},
         year = 2022,
        month = sep,
       volume = {936},
       number = {2},
          eid = {165},
        pages = {165},
          doi = {10.3847/1538-4357/ac887d},
archivePrefix = {arXiv},
       eprint = {2110.04314},
 primaryClass = {astro-ph.GA},
       adsurl = {https://ui.adsabs.harvard.edu/abs/2022ApJ...936..165L}
}

@ARTICLE{2023ApJ...951..137K,
       author = {{Kim}, Juhan and {Lee}, Jaehyun and {Laigle}, Clotilde and {Dubois}, Yohan and {Kim}, Yonghwi and {Park}, Changbom and {Pichon}, Christophe and {Gibson}, Brad K. and {Few}, C. Gareth and {Shin}, Jihye and {Snaith}, Owain},
        title = "{Low-surface-brightness Galaxies are Missing in the Observed Stellar Mass Function}",
      journal = {\apj},
         year = 2023,
        month = jul,
       volume = {951},
       number = {2},
          eid = {137},
        pages = {137},
          doi = {10.3847/1538-4357/acd251},
archivePrefix = {arXiv},
       eprint = {2212.14539},
 primaryClass = {astro-ph.GA},
       adsurl = {https://ui.adsabs.harvard.edu/abs/2023ApJ...951..137K}
}

@ARTICLE{2016ApJS..224...24L,
       author = {{Laigle}, C. and {McCracken}, H.~J. and {Ilbert}, O. and {Hsieh}, B.~C. and {Davidzon}, I. and {Capak}, P. and {Hasinger}, G. and {Silverman}, J.~D. and {Pichon}, C. and {Coupon}, J. and {Aussel}, H. and {Le Borgne}, D. and {Caputi}, K. and {Cassata}, P. and {Chang}, Y.-Y. and {Civano}, F. and {Dunlop}, J. and {Fynbo}, J. and {Kartaltepe}, J.~S. and {Koekemoer}, A. and {Le F{\`e}vre}, O. and {Le Floc'h}, E. and {Leauthaud}, A. and {Lilly}, S. and {Lin}, L. and {Marchesi}, S. and {Milvang-Jensen}, B. and {Salvato}, M. and {Sanders}, D.~B. and {Scoville}, N. and {Smolcic}, V. and {Stockmann}, M. and {Taniguchi}, Y. and {Tasca}, L. and {Toft}, S. and {Vaccari}, Mattia and {Zabl}, J.},
        title = "{The COSMOS2015 Catalog: Exploring the 1 < z < 6 Universe with Half a Million Galaxies}",
      journal = {\apjs},
         year = 2016,
        month = jun,
       volume = {224},
       number = {2},
          eid = {24},
        pages = {24},
          doi = {10.3847/0067-0049/224/2/24},
archivePrefix = {arXiv},
       eprint = {1604.02350},
 primaryClass = {astro-ph.GA},
       adsurl = {https://ui.adsabs.harvard.edu/abs/2016ApJS..224...24L}
}

@ARTICLE{2014ApJS..214...24S,
       author = {{Skelton}, Rosalind E. and {Whitaker}, Katherine E. and {Momcheva}, Ivelina G. and {Brammer}, Gabriel B. and {van Dokkum}, Pieter G. and {Labb{\'e}}, Ivo and {Franx}, Marijn and {van der Wel}, Arjen and {Bezanson}, Rachel and {Da Cunha}, Elisabete and {Fumagalli}, Mattia and {F{\"o}rster Schreiber}, Natascha and {Kriek}, Mariska and {Leja}, Joel and {Lundgren}, Britt F. and {Magee}, Daniel and {Marchesini}, Danilo and {Maseda}, Michael V. and {Nelson}, Erica J. and {Oesch}, Pascal and {Pacifici}, Camilla and {Patel}, Shannon G. and {Price}, Sedona and {Rix}, Hans-Walter and {Tal}, Tomer and {Wake}, David A. and {Wuyts}, Stijn},
        title = "{3D-HST WFC3-selected Photometric Catalogs in the Five CANDELS/3D-HST Fields: Photometry, Photometric Redshifts, and Stellar Masses}",
      journal = {\apjs},
         year = 2014,
        month = oct,
       volume = {214},
       number = {2},
          eid = {24},
        pages = {24},
          doi = {10.1088/0067-0049/214/2/24},
archivePrefix = {arXiv},
       eprint = {1403.3689},
 primaryClass = {astro-ph.GA},
       adsurl = {https://ui.adsabs.harvard.edu/abs/2014ApJS..214...24S}
}

@article{james_measurement_2023,
       author = {{James}, C.~W. and {Ghosh}, E.~M. and {Prochaska}, J.~X. and {Bannister}, K.~W. and {Bhandari}, S. and {Day}, C.~K. and {Deller}, A.~T. and {Glowacki}, M. and {Gordon}, A.~C. and {Heintz}, K.~E. and {Marnoch}, L. and {Ryder}, S.~D. and {Scott}, D.~R. and {Shannon}, R.~M. and {Tejos}, N.},
        title = "{A measurement of Hubble's Constant using Fast Radio Bursts}",
      journal = {\mnras},
         year = 2022,
        month = nov,
       volume = {516},
       number = {4},
        pages = {4862-4881},
          doi = {10.1093/mnras/stac2524},
archivePrefix = {arXiv},
       eprint = {2208.00819},
 primaryClass = {astro-ph.CO},
       adsurl = {https://ui.adsabs.harvard.edu/abs/2022MNRAS.516.4862J}
}

@article{chimefrb_collaboration_bright_2020,
       author = {{CHIME/FRB Collaboration} and {Andersen}, B.~C. and {Bandura}, K.~M. and {Bhardwaj}, M. and {Bij}, A. and {Boyce}, M.~M. and {Boyle}, P.~J. and {Brar}, C. and {Cassanelli}, T. and {Chawla}, P. and {Chen}, T. and {Cliche}, J.-F. and {Cook}, A. and {Cubranic}, D. and {Curtin}, A.~P. and {Denman}, N.~T. and {Dobbs}, M. and {Dong}, F.~Q. and {Fandino}, M. and {Fonseca}, E. and {Gaensler}, B.~M. and {Giri}, U. and {Good}, D.~C. and {Halpern}, M. and {Hill}, A.~S. and {Hinshaw}, G.~F. and {H{\"o}fer}, C. and {Josephy}, A. and {Kania}, J.~W. and {Kaspi}, V.~M. and {Landecker}, T.~L. and {Leung}, C. and {Li}, D.~Z. and {Lin}, H.-H. and {Masui}, K.~W. and {McKinven}, R. and {Mena-Parra}, J. and {Merryfield}, M. and {Meyers}, B.~W. and {Michilli}, D. and {Milutinovic}, N. and {Mirhosseini}, A. and {M{\"u}nchmeyer}, M. and {Naidu}, A. and {Newburgh}, L.~B. and {Ng}, C. and {Patel}, C. and {Pen}, U.-L. and {Pinsonneault-Marotte}, T. and {Pleunis}, Z. and {Quine}, B.~M. and {Rafiei-Ravandi}, M. and {Rahman}, M. and {Ransom}, S.~M. and {Renard}, A. and {Sanghavi}, P. and {Scholz}, P. and {Shaw}, J.~R. and {Shin}, K. and {Siegel}, S.~R. and {Singh}, S. and {Smegal}, R.~J. and {Smith}, K.~M. and {Stairs}, I.~H. and {Tan}, C.~M. and {Tendulkar}, S.~P. and {Tretyakov}, I. and {Vanderlinde}, K. and {Wang}, H. and {Wulf}, D. and {Zwaniga}, A.~V.},
        title = "{A bright millisecond-duration radio burst from a Galactic magnetar}",
      journal = {\nat},
         year = 2020,
        month = nov,
       volume = {587},
       number = {7832},
        pages = {54-58},
          doi = {10.1038/s41586-020-2863-y},
archivePrefix = {arXiv},
       eprint = {2005.10324},
 primaryClass = {astro-ph.HE},
       adsurl = {https://ui.adsabs.harvard.edu/abs/2020Natur.587...54C}
}

@ARTICLE{Driver2016,
       author = {{Driver}, Simon P. and {Andrews}, Stephen K. and {Davies}, Luke J. and {Robotham}, Aaron S.~G. and {Wright}, Angus H. and {Windhorst}, Rogier A. and {Cohen}, Seth and {Emig}, Kim and {Jansen}, Rolf A. and {Dunne}, Loretta},
        title = "{Measurements of Extragalactic Background Light from the Far UV to the Far IR from Deep Ground- and Space-based Galaxy Counts}",
      journal = {\apj},
         year = 2016,
        month = aug,
       volume = {827},
       number = {2},
          eid = {108},
        pages = {108},
          doi = {10.3847/0004-637X/827/2/108},
archivePrefix = {arXiv},
       eprint = {1605.01523},
 primaryClass = {astro-ph.GA},
       adsurl = {https://ui.adsabs.harvard.edu/abs/2016ApJ...827..108D}
}

@INPROCEEDINGS{2020ASPC..527..461B,
       author = {{Bertin}, E. and {Schefer}, M. and {Apostolakos}, N. and {{\'A}lvarez-Ayll{\'o}n}, A. and {Dubath}, P. and {K{\"u}mmel}, M.},
        title = "{The SourceXtractor++ Software}",
    booktitle = {Astronomical Data Analysis Software and Systems XXIX},
         year = 2020,
       editor = {{Pizzo}, R. and {Deul}, E.~R. and {Mol}, J.~D. and {de Plaa}, J. and {Verkouter}, H.},
       series = {Astronomical Society of the Pacific Conference Series},
       volume = {527},
        month = jan,
        pages = {461},
       adsurl = {https://ui.adsabs.harvard.edu/abs/2020ASPC..527..461B}
}

@INPROCEEDINGS{2020ASPC..527...29K,
       author = {{K{\"u}mmel}, M. and {Bertin}, E. and {Schefer}, M. and {Apostolakos}, N. and {{\'A}lvarez-Ayll{\'o}n}, A. and {Dubath}, P.},
        title = "{Working With the SourceXtractor++ Software}",
    booktitle = {Astronomical Data Analysis Software and Systems XXIX},
         year = 2020,
       editor = {{Pizzo}, R. and {Deul}, E.~R. and {Mol}, J.~D. and {de Plaa}, J. and {Verkouter}, H.},
       series = {Astronomical Society of the Pacific Conference Series},
       volume = {527},
        month = jan,
        pages = {29},
       adsurl = {https://ui.adsabs.harvard.edu/abs/2020ASPC..527...29K}
}

@ARTICLE{dlis,
       author = {{Dey}, Arjun and {Schlegel}, David J. and {Lang}, Dustin and {Blum}, Robert and {Burleigh}, Kaylan and {Fan}, Xiaohui and {Findlay}, Joseph R. and {Finkbeiner}, Doug and {Herrera}, David and {Juneau}, St{\'e}phanie and others},
        title = "{Overview of the DESI Legacy Imaging Surveys}",
      journal = {\aj},
         year = 2019,
        month = may,
       volume = {157},
       number = {5},
          eid = {168},
        pages = {168},
          doi = {10.3847/1538-3881/ab089d},
archivePrefix = {arXiv},
       eprint = {1804.08657},
 primaryClass = {astro-ph.IM},
       adsurl = {https://ui.adsabs.harvard.edu/abs/2019AJ....157..168D}
}

@ARTICLE{ps1,
       author = {{Chambers}, K.~C. and {Magnier}, E.~A. and {Metcalfe}, N. and {Flewelling}, H.~A. and {Huber}, M.~E. and {Waters}, C.~Z. and {Denneau}, L. and {Draper}, P.~W. and {Farrow}, D. and {Finkbeiner}, D.~P. and others},
        title = "{The Pan-STARRS1 Surveys}",
      journal = {arXiv e-prints},
         year = 2016,
        month = dec,
          eid = {arXiv:1612.05560},
        pages = {arXiv:1612.05560},
          doi = {10.48550/arXiv.1612.05560},
archivePrefix = {arXiv},
       eprint = {1612.05560},
 primaryClass = {astro-ph.IM},
       adsurl = {https://ui.adsabs.harvard.edu/abs/2016arXiv161205560C}
}

@ARTICLE{kolmogorov,
	author = {{Kolmogorov}, A.},
    title = {Sulla determinazione empirica di una legge di distribuzione},
  journal = {G.\ Ist.\ Ital.\ Attuari.},
     year = 1933,
   volume = 4,
    pages = {83-91}
}

@ARTICLE{smirnov,
	author = {{Smirnov}, N.},
    title = {Table for estimating the goodness of fit of empirical distributions},
  journal = {Annals of Mathematical Statistics},
     year = 1948,
   volume = 19,
    pages = {279-281},
    doi = {doi:10.1214/aoms/1177730256}
}

@ARTICLE{2026SciBu..71...76N,
       author = {{Niu}, Chen-Hui and {Li}, Di and {Yang}, Yuan-Pei and {Zhu}, Yuhao and {Zhang}, Yongkun and {Zhang}, Jia-Heng and {Du}, Zexin and {Yao}, Jumei and {Zheng}, Xiaoping and {Wang}, Pei and {Feng}, Yi and {Zhang}, Bing and {Zhu}, Weiwei and {Yu}, Wenfei and {Jiang}, Ji-An and {Dai}, Shi and {Tsai}, Chao-Wei and {Chen}, A. Ming and {Hou}, Yijun and {Niu}, Jiarui and {Wang}, Weiyang and {Miao}, Chenchen and {Li}, Xinming and {Zhang}, Junshuo},
        title = "{A persistently active fast radio burst source embedded in an expanding supernova remnant}",
      journal = {Science Bulletin},
         year = 2026,
        month = jan,
       volume = {71},
       number = {1},
        pages = {76-82},
          doi = {10.1016/j.scib.2025.11.023},
archivePrefix = {arXiv},
       eprint = {2512.05448},
 primaryClass = {astro-ph.HE},
       adsurl = {https://ui.adsabs.harvard.edu/abs/2026SciBu..71...76N}
}

@ARTICLE{2026ApJ...996L..16M,
       author = {{Moroianu}, Alexandra M. and {Bhandari}, Shivani and {Drout}, Maria R. and {Hessels}, Jason W.~T. and {Hewitt}, Dant{\'e} M. and {Kirsten}, Franz and {Marcote}, Benito and {Pleunis}, Ziggy and {Snelders}, Mark P. and {Sridhar}, Navin and {Bach}, Uwe and {Bempong-Manful}, Emmanuel K. and {Bezrukovs}, Vladislavs and {Blaauw}, Richard and {Bray}, Justin D. and {Buttaccio}, Salvatore and {Chatterjee}, Shami and {Corongiu}, Alessandro and {Feiler}, Roman and {Gaensler}, B.~M. and {Gawro{\'n}ski}, Marcin P. and {Giroletti}, Marcello and {Ibik}, Adaeze L. and {Karuppusamy}, Ramesh and {Lazda}, Mattias and {Leung}, Calvin and {Lindqvist}, Michael and {Masui}, Kiyoshi W. and {Michilli}, Daniele and {Nimmo}, Kenzie and {Ould-Boukattine}, Omar S. and {Pandhi}, Ayush and {Paragi}, Zsolt and {Pearlman}, Aaron B. and {Puchalska}, Weronika and {Scholz}, Paul and {Shin}, Kaitlyn and {Sluman}, Jurjen J. and {Trudu}, Matteo and {Williams-Baldwin}, David and {Yang}, Jun},
        title = "{A Milliarcsecond Localization Associates FRB 20190417A with a Compact Persistent Radio Source and an Extreme Magnetoionic Environment}",
      journal = {\apjl},
         year = 2026,
        month = jan,
       volume = {996},
       number = {1},
          eid = {L16},
        pages = {L16},
          doi = {10.3847/2041-8213/ae28c7},
archivePrefix = {arXiv},
       eprint = {2509.05174},
 primaryClass = {astro-ph.HE},
       adsurl = {https://ui.adsabs.harvard.edu/abs/2026ApJ...996L..16M}
}

@ARTICLE{2026ApJ...997L...6Y,
       author = {{Yamasaki}, Shotaro and {Hashimoto}, Tetsuya and {Kusakabe}, Haruka and {Goto}, Tomotsugu},
        title = "{No Metallicity Preference in Fast Radio Burst Host Galaxies}",
      journal = {\apjl},
         year = 2026,
        month = jan,
       volume = {997},
       number = {1},
          eid = {L6},
        pages = {L6},
          doi = {10.3847/2041-8213/ae2d19},
archivePrefix = {arXiv},
       eprint = {2508.07688},
 primaryClass = {astro-ph.HE},
       adsurl = {https://ui.adsabs.harvard.edu/abs/2026ApJ...997L...6Y}
}

@ARTICLE{2020Natur.587...59B,
       author = {{Bochenek}, C.~D. and {Ravi}, V. and {Belov}, K.~V. and {Hallinan}, G. and {Kocz}, J. and {Kulkarni}, S.~R. and {McKenna}, D.~L.},
        title = "{A fast radio burst associated with a Galactic magnetar}",
      journal = {\nat},
         year = 2020,
        month = nov,
       volume = {587},
       number = {7832},
        pages = {59-62},
          doi = {10.1038/s41586-020-2872-x},
archivePrefix = {arXiv},
       eprint = {2005.10828},
 primaryClass = {astro-ph.HE},
       adsurl = {https://ui.adsabs.harvard.edu/abs/2020Natur.587...59B}
}

@ARTICLE{niu_repeating_2022,
       author = {{Niu}, C.-H. and {Aggarwal}, K. and {Li}, D. and {Zhang}, X. and {Chatterjee}, S. and {Tsai}, C. -W. and {Yu}, W. and {Law}, C.~J. and {Burke-Spolaor}, S. and {Cordes}, J.~M. and {Zhang}, Y. -K. and {Ocker}, S.~K. and {Yao}, J. -M. and {Wan}, P. and {Feng}, Y. and {Niino}, Y. and {Bochenek}, C. and {Cruces}, M. and {Connor}, L. and {Jiang}, J. -A. and {Dai}, S. and {Luo}, R. and {Li}, G. -D. and {Miao}, C. -C. and {Niu}, J. -R. and {Anna-Thomas}, R. and {Sydnor}, J. and {Stern}, D. and {Wang}, W. -Y. and {Yuan}, M. and {Yue}, Y. -L. and {Zhou}, D. -J. and {Yan}, Z. and {Zhu}, W. -W. and {Zhang}, B.},
        title = "{A repeating fast radio burst associated with a persistent radio source}",
      journal = {\nat},
         year = 2022,
        month = jun,
       volume = {606},
       number = {7916},
        pages = {873-877},
          doi = {10.1038/s41586-022-04755-5},
archivePrefix = {arXiv},
       eprint = {2110.07418},
 primaryClass = {astro-ph.HE},
       adsurl = {https://ui.adsabs.harvard.edu/abs/2022Natur.606..873N}
}

@ARTICLE{Connor2019,
       author = {{Connor}, Liam},
        title = "{Interpreting the distributions of FRB observables}",
      journal = {\mnras},
         year = 2019,
        month = aug,
       volume = {487},
       number = {4},
        pages = {5753-5763},
          doi = {10.1093/mnras/stz1666},
archivePrefix = {arXiv},
       eprint = {1905.00755},
 primaryClass = {astro-ph.HE},
       adsurl = {https://ui.adsabs.harvard.edu/abs/2019MNRAS.487.5753C}
}

@ARTICLE{2025NatAs...9.1226C,
       author = {{Connor}, Liam and {Ravi}, Vikram and {Sharma}, Kritti and {Ocker}, Stella Koch and {Faber}, Jakob and {Hallinan}, Gregg and {Harnach}, Charlie and {Hellbourg}, Greg and {Hobbs}, Rick and {Hodge}, David and {Hodges}, Mark and {Kosogorov}, Nikita and {Lamb}, James and {Law}, Casey and {Rasmussen}, Paul and {Sherman}, Myles and {Somalwar}, Jean and {Weinreb}, Sander and {Woody}, David and {Konietzka}, Ralf M.},
        title = "{A gas-rich cosmic web revealed by the partitioning of the missing baryons}",
      journal = {Nature Astronomy},
         year = 2025,
        month = aug,
       volume = {9},
        pages = {1226-1239},
          doi = {10.1038/s41550-025-02566-y},
archivePrefix = {arXiv},
       eprint = {2409.16952},
 primaryClass = {astro-ph.CO},
       adsurl = {https://ui.adsabs.harvard.edu/abs/2025NatAs...9.1226C}
}

@article{cordes_ne2001,
       author = {{Cordes}, J.~M. and {Lazio}, T.~J.~W.},
        title = "{NE2001. II. Using Radio Propagation Data to Construct a Model for the Galactic Distribution of Free Electrons}",
      journal = {arXiv e-prints},
         year = 2003,
        month = jan,
          eid = {astro-ph/0301598},
        pages = {astro-ph/0301598},
          doi = {10.48550/arXiv.astro-ph/0301598},
archivePrefix = {arXiv},
       eprint = {astro-ph/0301598},
 primaryClass = {astro-ph},
       adsurl = {https://ui.adsabs.harvard.edu/abs/2003astro.ph..1598C}
}

@article{lorimer_bright_2007,
       author = {{Lorimer}, D.~R. and {Bailes}, M. and {McLaughlin}, M.~A. and {Narkevic}, D.~J. and {Crawford}, F.},
        title = "{A Bright Millisecond Radio Burst of Extragalactic Origin}",
      journal = {Science},
         year = 2007,
        month = nov,
       volume = {318},
       number = {5851},
        pages = {777},
          doi = {10.1126/science.1147532},
archivePrefix = {arXiv},
       eprint = {0709.4301},
 primaryClass = {astro-ph},
       adsurl = {https://ui.adsabs.harvard.edu/abs/2007Sci...318..777L}
}
